\documentclass[]{aastex63}
\usepackage{graphicx}	
\usepackage{amsmath}	
\usepackage{amssymb}	
\usepackage{breqn}
\usepackage{braket}
\usepackage{chemfig}
\usepackage{float}
\usepackage{wasysym}
\usepackage{mathtools}
\usepackage{xcolor}
\usepackage{cleveref}
\usepackage{multirow}
\usepackage{hhline}

\submitjournal{ApJ}

\shorttitle{Modelling Triatomic Biosignatures: Ozone}
\shortauthors{Thomas M.\ Cross, David M.\ Benoit, Marco Pignatari and Brad K.\ Gibson}

\graphicspath{{./}{figures/}}

\begin{document}

\title{Modelling Triatomic Biosignatures:\\
Ozone and Isotopomers}

\correspondingauthor{Thomas M. Cross}
\email{T.Cross-2019@hull.ac.uk}

\author[0000-0001-6621-8182]{Thomas M.\ Cross}
\affiliation{E.~A.~Milne Centre for Astrophysics, Department of Physics and
                Mathematics, University of Hull, HU6 7RX, United Kingdom}
                
\author[0000-0002-7773-6863]{David M.\ Benoit}
\affiliation{E.~A.~Milne Centre for Astrophysics, Department of Physics and
                Mathematics, University of Hull, HU6 7RX, United Kingdom}

\author[0000-0002-9048-6010]{Marco Pignatari}
\affiliation{Konkoly Observatory, Research Centre for Astronomy and Earth Sciences, HUN-REN, \\ Konkoly Thege Mikl\'{o}s \'{u}t 15-17, H-1121 Budapest, Hungary}
\affiliation{CSFK, MTA Centre of Excellence, Budapest, Konkoly Thege Mikl\'{o}s \'{u}t 15-17, H-1121, Hungary}
\affiliation{E.~A.~Milne Centre for Astrophysics, Department of Physics and
                Mathematics, University of Hull, HU6 7RX, United Kingdom}
\affiliation{NuGrid Collaboration, \url{http://nugridstars.org}}
\affiliation{Joint Institute for Nuclear Astrophysics - Center for the Evolution of the Elements}

\author[0000-0003-4446-3130]{Brad K.\ Gibson}

\affiliation{Astrophysics Group, Keele University, Keele, Staffordshire ST5 5BG, United Kingdom}

\begin{abstract}

In this work we present a new approach to produce spectroscopic constants and model first-principles synthetic spectra for all molecules of astrophysical interest. We have generalized our previous diatomic molecule simulation framework, employing Transition-Optimised Shifted Hermite (TOSH) theory, thereby enabling the modelling of polyatomic rotational constants for molecules with three or more atoms. These capabilities, are now provided by our new code Epimetheus.
As a first validation of our approach, we confront our predictions and assess their accuracy against the well-studied triatomic molecule, ozone 666 ($^{16}$O$_3$), in addition to eight of its potential isotopomers: ozone 668 ($^{16}$O$^{16}$O$^{18}$O), 686 ($^{16}$O$^{18}$O$^{16}$O), 667 ($^{16}$O$^{16}$O$^{17}$O), 676 ($^{16}$O$^{17}$O$^{16}$O), 688 ($^{16}$O$^{18}$O$^{18}$O), 868 ($^{18}$O$^{16}$O$^{18}$O), 888 ($^{18}$O$_3$), and 777 ($^{17}$O$_3$). We then assess the accuracy of these rotational constants using the Epimetheus data in our code Pandora, and generate synthetic molecular spectra.
The ozone spectra presented here are purely infrared and not Raman. Epimetheus builds upon the work from our previous code Prometheus, which used the TOSH theory to account for anharmonicity for the fundamental $\nu=0 \rightarrow \nu=1$ band, going further to now account for triatomic molecules. 
This is combined with thermal profile modeling for the rotational transitions. 
We have found that this extended method performs promisingly, typically approximating the spectroscopic constants and spectra well. Some issues do arise depending on the symmetry group of the ozone isotopomer. In general, we show that Epimetheus can provide the data to produce appreciable molecular spectra, to help drive future high-resolution studies.

\end{abstract}


\section{Introduction}
\label{section:Introduction}

The work here provides an extension to the methodology described in our previous paper to model biosignatures \citep{22CrBePi}. Molecular lines are ideal observation targets to study different astrophysical bodies, such as exoplanetary atmospheres for biosignatures. However, apart from extremely well studied molecules, the basic data needed to detect most biosignatures in spectra is incomplete. We aim to now to extend past the diatomic molecules explored previously and produce spectroscopic constants (and to a lesser degree spectra) for polyatomic molecules. Specifically within this paper, we will be focusing first on the triatomic molecules, with an extension to larger polyatomic molecules in a future publication. Like before we are trying to address the methodological gap between approximate, fundamental-only models (such as harmonic,  e.g., \citet{21DuBeRi}), and labour intensive, extremely accurate, line lists (such as ExoMol and HITRAN, see e.g., \citet{20TeYuAl} and \citet{ 22GoRoHa}) . This is an area of research which has begun to garner interest, as the rapid production of vibrational spectral data is paramount for the identification of molecules in exoplanetary atmospheres \citep{23ZaJuPe}.

Generally a biosignature is defined as an object, substance, and/or pattern whose origin specifically requires a biological agent \citep{99MaWaXX}. In the context of astronomical observations however this is used to mean a gas that is produced by life and accumulates in a planet's atmosphere. An ideal biosignature would be unambiguous with living organisms being its unique source \citep[][]{20GrRiBa}, in reality though, many biosignatures can be also produced through abiotic processes and therefore can act as false positives \citep{18HaDoXX}. There is currently no globally accepted scheme for classifying what molecules are potentially atmospheric biosignatures \citep{18ScEdKi}, which means we have to select one we believe would be best. 

In this work we have opted to use a scheme by \citet{16SeBaPe}, who through a cross-discipline scientific effort, have created a list called "The All Small Molecules" catalogue (ASM), which contains over 16,000 potential biosignature molecules. Out of the potential pool of candidates within the ASM, we have selected ozone as our test case for this study. This was done for a variety of reasons: first, the fact that ozone is an asymmetric top (for more details see \citet{99MiBaTy}), and thus it differs from a linear triatomic molecule, such as carbon dioxide, which would effectively model like a simpler diatomic (see our previous approach in \citet{22CrBePi}). Additionally ozone has been thoroughly investigated and plenty of experimental data is available for us to benchmark our synthetic data \citep{97FlBaXX, 20GaTeKe, 01HaMaXX, 90FlCaRi, 22GoRoHa, 03FlPiCa, 02WaBiSc, 21TyBaMi, 18BaStBa, 91RiSmDe, 87RoGaGo, 85GaXXXX}.

It is known that, on Earth, ozone appears throughout the entire atmosphere but resides prominently in a layer in the lower stratosphere at an altitude between 15km and 30km \citep[][]{78PrAlCu}. Ozone is crucial for our atmosphere as it is responsible for the temperature inversion effect within the stratosphere \citep[][]{11KaXXXX}, where temperature is reliant upon the solar energy absorbed. This ultimately means this stratospheric ozone reduces the incoming ultraviolet radiation \citep[][]{99SoXXXX} to levels acceptable for not only life's development but also its continuation \citep[][]{11MoBeLe}.

Ozone is also important because of its relationship to another key biosignature - diatomic oxygen (O$_2$). On Earth O$_2$ is produced from oxygenic photosynthesis, whether it be via flora or fauna, by for example cyanobacteria \citep[][]{23BaCoCl}. Ozone is created photochemically in the atmosphere from O$_2$, consequently meaning its concentration depends heavily upon the production of O$_2$ \citep[][]{14DoSeCl}. The reactions which naturally form and destroy ozone, known as the Chapman reactions \citep[][]{30ChXXXX}, are described in the equations \ref{eq:chapmanchem_r1}, \ref{eq:chapmanchem_r2}, \ref{eq:chapmanchem_r3} and \ref{eq:chapmanchem_r7} below. Please note that species M in these equations denotes a third body required to carry away excess vibrational energy.

\begin{equation}
\label{eq:chapmanchem_r1}
    \textrm{O}_2 + h\nu \rightarrow 2\textrm{O}
\end{equation}
\begin{equation}
\label{eq:chapmanchem_r2}
    \textrm{O} + \textrm{O}_2 + \textrm{M} \rightarrow \textrm{O}_3 + \textrm{M}
\end{equation}
\begin{equation}
\label{eq:chapmanchem_r3}
    \textrm{O}_3 + h\nu \rightarrow \textrm{O}_2 + \textrm{O}
\end{equation}
\begin{equation}
\label{eq:chapmanchem_r7}
    \textrm{O}_3 + \textrm{O} \rightarrow \textrm{O}_2 + \textrm{O}_2
\end{equation}

This is not the only means of production of O$_2$, as it can be formed abiotically in atmospheres with high concentrations of carbon dioxide (CO$_2$) \citep[][]{07SeMeKa, 13GrGeGo}. 
Abiotic sources of ozone have been detected on two terrestrial planets within our solar system, Mars \citep[][]{09FaKoLe, olsen:20} and on Venus \citep[][]{11MoBeLe, marcq:19}, both of which are currently understood to be uninhabited by life. 
Further studies have shown that the ozone in the atmosphere of Mars has been produced from the photolysis of the high levels CO$_2$ \citep[][]{21LeTrFe} and then the Chapman reactions \citep[][]{30ChXXXX}, thus confirming an example of ozone's creation from abiotic means. Minor traces of ozone have also been detected on the Jupiter moon Ganymede \citep[][]{noll:96}.

From a physical chemistry standpoint, ozone is a bent asymmetric rotor. The ``symmetric" isotopic variants (such as $^{16}$O$_3$) belong to the $\rm{C_{2V}}$ point symmetry group, and the ``asymmetric" species (such as $^{16}$O$^{16}$O$^{18}$O) belong to the $\rm{C_S}$ point group \citep[][]{99MiBaTy}. This is important as the differences in symmetry point groups has an appreciable effect on the possible energetic transitions accessible to each species, in turn altering the spectra produced, and will be discussed further in Section \ref{subsection:generalSpecModel}.

\begin{figure}[ht]
    \centering
    \includegraphics[width=\columnwidth]{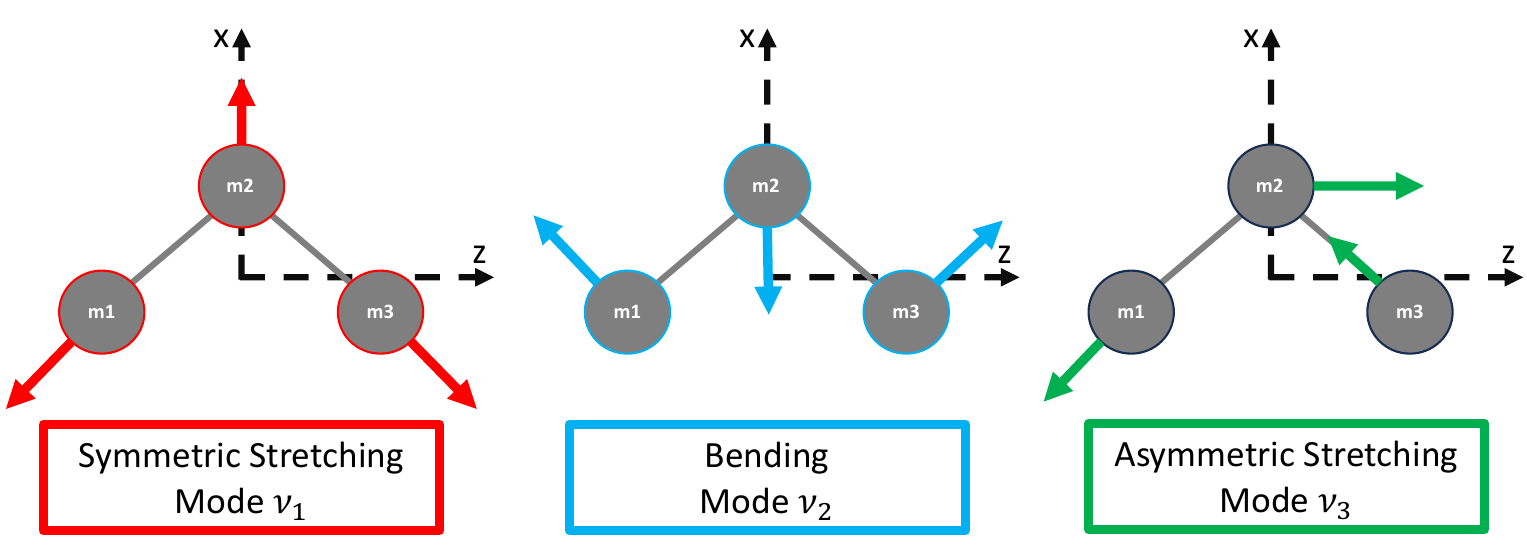}
    \caption{The three vibrational normal modes for a non-linear triatomic, such as ozone. }
    \label{fig:Introduction_Graphics/Ozone_fundamental_modes_diagram}
\end{figure}

The vibrational states of ozone are usually described using three quantum numbers, $n_1$, $n_2$ and $n_3$. These are often written simply as ``$n_1n_2n_3$". Each of these quantum numbers describe the degree of excitation of the three normal modes; the symmetric stretch ($\nu_1$), bend ($\nu_2$) and asymmetric stretch ($\nu_3$) \citep{72BaXXXX}. For example, the band ``001", describes when ozone has a single excitation in the asymmetric stretch vibration, $\nu_3$, but no excitation in the symmetric stretch and bending vibration. Bands which contain only a single excitation in one type of vibration are known as fundamental bands and are the focus of this paper. The normal modes of vibration for the fundamental bands of ozone (100, 010 and 001, in this case) are shown in Figure \ref{fig:Introduction_Graphics/Ozone_fundamental_modes_diagram}.

The isotopic variants for which we investigate band origins, rotational constants and spectra are shown in table \ref{tab:iso_abundance}. The order of these variants has been decided by their abundance, starting with the most abundant and then decreasing. The other isotopologues not shown in table \ref{tab:iso_abundance} are present within Section \ref{subsection:other_bandorigins}, but have only been evaluated by their band origins. For a full abundance list with all 18 stable isotopologues please refer to Table 2 of \citet{22BaMiSt}. The table numbers which correspond to the isotopologues spectroscopic constants has been included table \ref{tab:iso_abundance}, as well as the figure numbers for the corresponding spectra.

\begin{table}[ht]
    \centering
    \renewcommand{\arraystretch}{1.2}
    \begin{tabular}{|c|c|c|c|c|c|}
        \hline
        Isotopic Species & Designation & Spins of Atoms & Natural Abundance & Table No. & Spectra Figure/s \\
        \hline \hline
         $^{16}$O$_3$ & 666 & 0\,0\,0 & 0.992901 & 
         \ref{tab:constant_comparisons_666_and_668} & 
         \ref{fig:EPI_O3_666_13DaLo_HARMONIC_HITRAN2020_300K}, \ref{fig:EPI_O3_666_13DaLo_98FlBa_HITRAN2020_300K}, \ref{fig:EPI_O3_666_13DaLo_18PoZo_HITRAN_subplots_300K}
         \\
         \hline
         $^{16}$O$^{16}$O$^{18}$O & 668 & 0\,0\,0 & 3.98194 × 10$^{-3}$ & \ref{tab:constant_comparisons_666_and_668} & \ref{fig:EPI_O3_668_686_667_megamix} (A)
         \\
         \hline
         $^{16}$O$^{18}$O$^{16}$O & 686 &  0\,0\,0 & 1.99097 × 10$^{-3}$ & \ref{tab:constant_comparisons_686_and_667} & \ref{fig:EPI_O3_668_686_667_megamix} (B)
         \\
         \hline
         $^{16}$O$^{16}$O$^{17}$O & 667 &  0\,0\,$\frac{5}{2}$ & 7.40475 × 10$^{-4}$  & \ref{tab:constant_comparisons_686_and_667} & \ref{fig:EPI_O3_668_686_667_megamix} (C)
         \\
         \hline
         $^{16}$O$^{17}$O$^{16}$O & 676 &  0\,$\frac{5}{2}$\,0 & 3.70237 × 10$^{-4}$  & \ref{tab:constant_comparisons_676_and_688} & \ref{fig:EPI_O3_676_868_888_megamix} (A)
         \\
         \hline 
         $^{16}$O$^{18}$O$^{18}$O & 688 &  0\,0\,0 &  8.384576 × 10$^{-6}$  & \ref{tab:constant_comparisons_676_and_688} & -
         \\
         \hline
         $^{18}$O$^{16}$O$^{18}$O & 868 & 0\,0\,0 & 4.192288 × 10$^{-6}$ & \ref{tab:constant_comparisons_868_and_888} & \ref{fig:EPI_O3_676_868_888_megamix} (B)
         \\
         \hline
         $^{18}$O$_3$ & 888 &  0\,0\,0 & 8.615 × 10$^{-9}$ & \ref{tab:constant_comparisons_868_and_888} & \ref{fig:EPI_O3_676_868_888_megamix} (C)
         \\
         \hline 
         $^{17}$O$_3$ & 777 &  $\frac{5}{2}$\,$\frac{5}{2}$\,$\frac{5}{2}$ & 5.5 × 10$^{-11}$  & \ref{tab:constant_comparisons_777} & -
         \\
         \hline
    \end{tabular}
    \caption{A list of ozone isotopomers and their natural abundances, truncated to include only the variants included within the major data reported within this paper. All data has been obtained from \citet{22BaMiSt}. The designation column refers to the abbreviation that the isotopologue may be given throughout this paper.}
    \label{tab:iso_abundance}
\end{table}

In Section \ref{section:MethodologyA}, we introduce the methodology and theory required to derive the spectroscopic constants which is used within our code, Epimetheus. The last portion of the methodology, \ref{subsection:generalSpecModel}, gives a brief description of the method used to model the rovibrational spectra. A more detailed account of our spectral modeling approach (Pandora) will be published in a separate paper. 
In Section \ref{section:ResultsA} we discuss our results and we compare them with the literature.
Section \ref{section:ResultsB} presents generated ro-vibrational spectra using data obtained in section \ref{section:ResultsA}, with comparisons to well-established databases. 
The most significant features are summarized for each isotopomer. 
We conclude our findings in Section \ref{section:Conclusions}.

\section{Methodology}
\label{section:MethodologyA}

This Section details the 
novel approach implemented in our new code Epimetheus.
Compared to our previous paper \citet{22CrBePi}, we developed a more general methodology to obtain the anharmonically corrected spectroscopic constants for polyatomic molecules. 
This includes the band origin for each of the fundamental vibrational modes and the three rotational constants, $A$, $B$ and $C$ as opposed to the singular $B$ rotational constant for diatomic molecules \citep[][]{15FiXXXX}{}{}.

We also briefly discuss the process by which we model the spectroscopic constants, and also show the key equations required to produce a spectrum. As the spectral modelling is done using our other code - Pandora - and the prominent focus of our study relates to the anharmonic corrections of spectroscopic constants, we will not go into too much detail here. The in depth spectral modelling methodology will be explained in a later paper.

Finally we describe the computational details of our method. This ranges from the choices of inputted data to the choice of method used to calculate anharmonic correction variables.

\subsection{Determination of the Band Origins}
\label{subsection:Methodology_HessMatrix}

To calculate the anharmonic corrections, we first compute a Hessian matrix $\pmb{f_{\rm{CART}}}$ \citep[][]{99OcXXXX} which contains the second partial derivatives of the potential, $V$, with respect to displacement of the atoms in Cartesian (CART) coordinates:

\begin{equation}
\label{eq:hessian_CART}
    f_{\rm{CART}ij} = \left(\frac{\partial^2V}{\partial \xi_i \partial\xi_j}\right)_0
\end{equation}

The matrix is $3N \times 3N$ in size, where $N$ is the number of atoms within a molecule. $\xi_1, \xi_2, \xi_3, \cdots \xi_{3N}$ are used to represent the displacement in Cartesian coordinates, $\Delta x_1, \Delta y_1, \Delta z_1, \cdots \Delta z_N$. The $()_0$ is used to indicate that the derivatives are taken at the equilibrium positions of the atoms and therefore the first derivatives are assumed to be zero. The matrix created by the Hessian is shown below in eq. \ref{eq:hessian_CART_expanded}.

\begin{equation}
\label{eq:hessian_CART_expanded}
    \pmb{f_{\rm{CART}}} =  
\left(\begin{matrix}
\frac{\partial^2V}{\partial \xi_1 \partial\xi_1} & \frac{\partial^2V}{\partial \xi_1 \partial\xi_2} & \frac{\partial^2V}{\partial \xi_1 \partial\xi_3} & \cdots &\frac{\partial^2V}{\partial \xi_1 \partial\xi_n} 
\\[6pt]
\frac{\partial^2V}{\partial \xi_2 \partial\xi_1} & \frac{\partial^2V}{\partial \xi_2 \partial\xi_2} & \frac{\partial^2V}{\partial \xi_2 \partial\xi_3} & \cdots &\frac{\partial^2V}{\partial \xi_2 \partial\xi_n} 
\\[6pt]
\vdots & \vdots & \vdots & \ddots & \vdots 
\\[6pt]
\frac{\partial^2V}{\partial \xi_n \partial\xi_1} & \frac{\partial^2V}{\partial \xi_n \partial\xi_2} & \frac{\partial^2V}{\partial \xi_n \partial\xi_3} & \cdots &\frac{\partial^2V}{\partial \xi_n \partial\xi_n} 
\end{matrix}
\right)
\end{equation}{}

In our implementation, we use the method derived by \citet{08LiGiGI} in their TOSH theory to compute the Hessian matrix. The equations to describe the elements under this framework have been included in Section \ref{subsection:Computational Details}.

The next step is to appropriately mass-weight the calculated second derivatives to convert them into mass-weighted coordinates (MWC).
\begin{equation}
\label{eq:hessian_MWC}
    f_{\rm{MWC}ij} = \frac{f_{\rm{CART}ij}}{\sqrt{m_i m_j}} = \left(\frac{\partial^2V}{\partial q_i \partial q_j}\right)_0
\end{equation}

Where $q_i\,=\, \sqrt{m_i}\xi_i$ is the mass weighted Cartesian coordinate for $\xi_i$. The mass-weighted coordinate (MWC) Hessian is the following:
\begin{equation}
\label{eq:hessian_MWC_expanded}
    \pmb{f_{\rm{MWC}}} =  \left(
\begin{matrix}
\frac{\partial^2V}{\partial q_1 \partial q_1} & \frac{\partial^2V}{\partial q_1 \partial q_2} & \frac{\partial^2V}{\partial q_1 \partial q_3} & \cdots &\frac{\partial^2V}{\partial q_1 \partial q_9}
\\[6pt]
\frac{\partial^2V}{\partial q_2 \partial q_1} & \frac{\partial^2V}{\partial q_2 \partial q_2} & \frac{\partial^2V}{\partial q_2 \partial q_3} & \cdots &\frac{\partial^2V}{\partial q_2 \partial q_9}
\\[6pt]
\vdots & \vdots & \vdots & \ddots & \vdots 
\\[6pt]
\frac{\partial^2V}{\partial q_9 \partial q_1} & \frac{\partial^2V}{\partial q_9 \partial q_2} & \frac{\partial^2V}{\partial q_9 \partial q_3} & \cdots &\frac{\partial^2V}{\partial q_9 \partial q_9}
\end{matrix}\right)
\end{equation}{}

Once the MWC Hessian has been obtained, the normal modes, $\pmb{Q}_i$, (eigenvectors) and corresponding harmonic vibrational frequencies, $\omega_i$, (eigenvalues) are obtained by diagonalising $\pmb{f_{\rm{MWC}}}$.
Since the lowest eigenvectors will correspond to global translations or rotations of the molecule (6 modes for non-linear molecule and 5 for a linear molecule), we only consider the $M=3N-6$ or $M=3N-5$ eigenvectors associated with the largest eigenvalues. Those eigenvalues can be converted in wave numbers using the conversion described in eq. \ref{eq:nu_eigenvalues_to_freq} below.
\begin{equation}
\label{eq:nu_eigenvalues_to_freq}
    \omega_i = \sqrt{\frac{\eta_{ii}}{4\pi^2c_{cm}^2}}
\end{equation}

The constant $c_{cm}$ is the speed of light  and $\eta_{ii}$ is the eigenvalue and $\omega$ is the harmonic frequency of mode $i$. To be able to obtain $\omega$ in wavenumbers ($\left[\textrm{cm}\right]^{-1}$), $c_{cm}$ must have the units of $\left[\textrm{cm}\right]^{-2}\left[\textrm{s}\right]^{-1}$ and $\eta_{ii}$ must be in the units of $\left[\textrm{J}\right]\left[\textrm{m}\right]^{-2}\left[\textrm{kg}\right]^{-1}$.

We use equation \ref{eq:nu_eigenvalues_to_freq} to computed the harmonic frequencies for each of the fundamental modes. The next step, detailed in Section \ref{subsection:Methodology_AnharmBand}, is to calculate further spectroscopic constants required to include anharmonic correction for those frequencies.

\subsection{Anharmonically Correcting the Band Origins}
\label{subsection:Methodology_AnharmBand}

We will now detail the process of correcting the band origins that have been calculated by the eigenvalues in eq. \ref{eq:hessian_MWC}. \citet{08LiGiGI} stated that for polyatomic molecules, the TOSH fundamental transition energy for the $i$th mode is described as:
\begin{equation}
\label{eq:Delta_E_TOSH}
    \Delta E_i^{TOSH} = \omega_i + \frac{1}{8\omega_i} \sum_j^M \frac{\eta_{iijj}}{\omega_j} + \frac{1}{2\omega_i} \sum_j^M \eta_{iij}\sigma_{ij} + \frac{1}{4\omega_i} \sum_{j}^M \eta_{iijj}\sigma_{ij}^2
\end{equation}

Where $\eta_{iijj}$ is the quartic constant, $\eta_{iij}$ is the cubic constant and $\sigma_{ij}$ is a shift parameter (see below and also \citet{08LiGiGI}). The method we use to compute the cubic and quartic constants is detailed in the computational Section \ref{subsection:Computational Details}.

A key term in this equation is 
$\sigma$, which is the parameter 
providing the shift to the harmonic oscillator functions from their equilibrium position, thus providing anharmonic corrections. The equation to calculate $\sigma$ differs from the one shown in our previous paper \citet{22CrBePi}, since we now need a $\sigma$ that can accurately represent the multi-mode nature of a polyatomic molecule as opposed to a single mode diatomic.

$\sigma$ is calculated by comparing eq. \ref{eq:Delta_E_TOSH} with the second order perturbation theory (VPT2) equation for the polyatomic fundamental energy difference. It is described as the following: 
\begin{equation}
\label{eq:sigma_ij}
    \sigma_{ij} = \frac{\left(\delta_{ij} - 2\right)\left(\omega_i + \omega_j\right)\eta_{iij}}{4\omega_i \omega_j^2 \left(2\omega_i + \omega_j \right)} - \sum_k^M \frac{\eta_{kkj}}{4\omega_k \omega_j^2}
\end{equation}

The variable $\delta_{ij}$ in this equation represents the Kronecker delta, where $\delta_{ij} = 0$ occurs when $i \neq j$ and when $i = j$ it becomes $\delta_{ij} = 1$. The notation here of $\eta_{kkj}$ does not imply $k \neq i$ in the summation component of eq. \ref{eq:sigma_ij}. The $\sigma_{ij}$ elements define the $\pmb{\sigma}$ matrix which has dimensions of $M\times M$.

\subsection{Obtaining the Polyatomic Rotational Constants}
\label{subsection:Methodology_RevRotConst}

Before we begin the anharmonic corrections to the equilibrium geometries we first need to calculate geometries with respect to the centre of mass (COM). The first step is to actually calculate the COM, this is done using the typical equation:
\begin{equation}
\label{eq:R_COM}
    \pmb{R}_{COM} = \frac{\sum_i^N m_i \pmb{r}_{i}}{\sum_i^N m_i}
\end{equation}

Where $\pmb{r}_{i}$ represents the equilibrium coordinates for the i$^{th}$ atom in a molecule. We can see from equation \ref{eq:R_COM} the coordinates need to be appropriately mass weighted to allow COM calculations. Then by taking the equilibrium position $\pmb{r}$ and subtracting the COM we can find the geometries of the molecule with respect to the COM.
\begin{equation}
\label{eq:r_COM}
    \pmb{r}_{COM} = \pmb{r} - \pmb{R}_{COM}
\end{equation}

For a complete example workout see the appendix Section \ref{ap:RCOM_workout}, where we run through how the centre of mass origin is done for an imaginary molecule ($\alpha$ $\beta$ $\gamma$). $\alpha$ $\beta$ $\gamma$ has been selected for a work-through rather than O$_3$ as it is easier to differentiate between the atoms.

We then calculate moments of inertia (diagonal elements) and the products of inertia (non-diagonal elements) of the moment of inertia tensor $\pmb{I}$. This is represented by eq. \ref{eq:Inertia_matrix} below. For a complete breakdown on how to calculate each component within the  moment of inertia tensor $\pmb{I}$, such as $I_{xx}$, of the inertia tensor please refer to the appendix, Section \ref{ap:InertiaParts}. 
\begin{equation}
\label{eq:Inertia_matrix}
    \pmb{I} =  \left(
\begin{matrix}
I_{xx} & I_{xy} & I_{xz}
\\[6pt]
I_{yx} & I_{yy} & I_{yz}
\\[6pt]
I_{zx} & I_{zy} & I_{zz}
\end{matrix}\right)
\end{equation}

The inertia matrix can be diagonalised into its corresponding eigenvalues (also known as the principal moments of inertia) and eigenvectors. 
As there are three eigenvalues, we will obtain 3 rotational constants; $A$, $B$ and $C$. These have the following relations to define them by:
\begin{equation}
   A = \frac{\hbar}{4\pi c_{cm} I_{A}} > B = \frac{\hbar}{4\pi c_{cm} I_{B}} > C = \frac{\hbar}{4\pi c_{cm} I_{C}} \quad \therefore \quad I_{C} > I_{B} > I_{A}
\label{eq:rotational_constant_relations}
\end{equation}

Like eq. \ref{eq:nu_eigenvalues_to_freq}, $c_{cm}$ represents the speed of light given in the appropriate units ($\left[\textrm{cm}\right]\left[\textrm{s}\right]^{-1}$) and $\hbar$ represents the reduced Planck constant.  Once again awareness need to be exercised with regards to the units as the eigenvalues will need to be in the correct unit form ($\left[\textrm{m}\right]^2\left[\textrm{kg}\right]$), for eq. \ref{eq:rotational_constant_relations} to produce rotational constants in the form of wave numbers ([$\textrm{cm}^{-1}$]).

\subsection{Anharmonically Correcting the Polyatomic Rotational Constants}
\label{subsection:Methodology_AnharmRevRotConst}
Similarly to our previous work \citep{22CrBePi}, we assume that the harmonic shift matrix $\pmb{\sigma}$, defined by Equation \ref{eq:sigma_ij}, captures the geometric changes caused by anharmonicity. Indeed, we showed for diatomics that shifting the equilibrium geometry by $\sigma$ (in that case a scalar) led to a good approximation of the vibrationally averaged geometry for the first excited vibrational state for diatomics. The shifted geometry enabled us to determine $B_1$ for that excited state to less than a percent in most cases. We present here a generalisation of this approach to all polyatomic molecules. 

First, to simplify the notation, we define an excitation vector as:
\begin{eqnarray}
    \pmb{n}=\left(\begin{matrix} 
    n_1\\
    n_2\\
    \vdots\\
    n_M
    \end{matrix}\right)
\end{eqnarray}
where $n_i$ represents the excitation quanta in mode $i$. For example, the ground state is defined as an M-dimensional zero vector, $\pmb{0}$. 
To compute the geometry distortion caused by a single excitation in mode $j$, denoted by excitation vector $\pmb{n}_j$, we compute the matrix-vector product:
\begin{eqnarray}
\pmb{\sigma} \pmb{n}_j = \left(\begin{matrix} 
\sigma_{1j}\\
\sigma_{2j}\\
    \vdots\\
\sigma_{Mj}
    \end{matrix}\right)
    \label{eq:sigmavec}
\end{eqnarray}
Here we assume that only mode $j$ is excited and all other modes are in their ground state. The matrix-vector product expression remains valid for other excitation patterns but the resulting vector will have a more complex expression. To further clarify the excitation pattern used, we index the variables with $0\ldots n_j\ldots0$ to represent the chosen excitation.

This resulting shift vector is expressed in the basis of normal-modes vectors determined earlier in Section \ref{subsection:Methodology_HessMatrix} and therefore the corresponding Cartesian displacement, $\pmb{\sigma}_{0\cdots n_j\cdots 0}$, can be written as:
\begin{eqnarray}
\pmb{\sigma}_{0\cdots n_j\cdots 0}=\sigma_{1j}\, \pmb{Q}_1 + \sigma_{2j}\, \pmb{Q}_2 + \cdots + \sigma_{Mj}\, \pmb{Q}_M
\end{eqnarray}
where the $\pmb{Q}_i$ are the Cartesian normal-mode eigenvectors obtained in Section \ref{subsection:Methodology_HessMatrix}. 

Finally, the displaced Cartesian geometry for a single excitation in mode $n_j$ is computed using the equilibrium Cartesian geometry $\pmb{X}_0$ as:
\begin{eqnarray}
\pmb{X}_{0\cdots n_j\cdots 0}=\pmb{X}_0+\pmb{\sigma}_{0\cdots n_j\cdots 0}
\end{eqnarray}
Using this new displaced nuclear configuration, we can obtain new rotational constants by re-computing the inertia tensor and following the procedure outlined earlier in Section \ref{subsection:Methodology_RevRotConst}. The entire process described then needs to be repeated for every modal excitation. 

Note in passing that, since the ground-state excitation vector is a zero vector (i.e.\ $\pmb{\sigma}_{0\cdots 0}=\pmb{0}$), the geometry used for the ground state rotational constants is the equilibrium Cartesian geometry $\pmb{X}_0$, as suggested for diatomic molecules in \citet{22CrBePi}.

In the particular case of ozone with 3 normal modes ($M=3$), the $\pmb{\sigma}$ matrix is of size $3\times 3$. The displaced geometry for ``mode 1" (denoted ``100'') can be obtained using the simple excitation vector $\pmb{n}$ below:
\begin{eqnarray}
    \pmb{n}=\left(\begin{matrix} 
    1\\
    0\\
    0
    \end{matrix}\right)
\end{eqnarray}
along with the expression \ref{eq:sigmavec} leading to:
\begin{eqnarray}
\pmb{\sigma}\pmb{n}_1=\left(\begin{matrix} 
\sigma_{11}\\
\sigma_{21}\\
\sigma_{31}
    \end{matrix}\right)
\end{eqnarray}
The result can be transformed into a Cartesian displacement vector, $\pmb{\sigma}_{100}$, as described earlier:
\begin{equation}
\pmb{\sigma}_{100}=\sigma_{11}\, \pmb{Q}_1 + \sigma_{21}\, \pmb{Q}_2 + \sigma_{31}\, \pmb{Q}_3
\end{equation}
Combining this shift with the equilibrium geometry, $\pmb{X}_0$:
\begin{equation}
\pmb{X}_{100} = \pmb{X}_0 + \pmb{\sigma}_{100}
\end{equation}
Where $\pmb{X}_{100}$ provides our estimate of the displaced Cartesian geometry for ozone in the 100 vibrational excited state used to compute the corresponding rotational constants, $A_{100}$, $B_{100}$ and $C_{100}$.   

\subsection{Spectral Modelling Overview}
\label{subsection:generalSpecModel}

In this Section we describe the process of modelling the spectra used with our other code Pandora. This ranges from deriving the possible transitions to the intensities of those transitions. We will only cover the essentials here and provide useful references for the reader to investigate should they wish to do so.

For the infrared rotational spectrum, provided a permanent dipole moment is present \citep[][]{45HeXXXX}, the selection rule for the J quantum number is $J = 0, \pm 1$. Therefore only the P, Q and R branches equations are needed. It is worth noticing that the following branch equations have been derived to accurately describe ozone, which is an asymmetric top molecule and therefore will not accurately describe the transitions for other types of molecules, such as a linear polyatomics which are not within the scope of this paper. Rotational energy levels of asymmetric tops can be described by the following three equations:
\begin{equation}
\label{eq:P_branch}
\bar{\nu}_{P} = \Delta E_i^{TOSH} + \frac{1}{2}\left(A_1 + C_1\right)J\left(J + 1\right) + \frac{1}{2}\left(A_1 - C_1\right) E_{\tau'} -\frac{1}{2}\left(A_0 + C_0\right)\left(J + 1\right)\left(J + 2\right) - \frac{1}{2}\left(A_0 - C_0\right) E_{\tau''}
\end{equation}

\begin{equation}
\label{eq:Q_branch}
\bar{\nu}_{Q} = \Delta E_i^{TOSH} + \frac{1}{2}\left(A_1 + C_1\right)J\left(J + 1\right) + \frac{1}{2}\left(A_1 - C_1\right) E_{\tau'} - \frac{1}{2}\left(A_0 + C_0\right)J\left(J + 1\right) - \frac{1}{2}\left(A_0 - C_0\right)E_{\tau''}
\end{equation}

\begin{equation}
\label{eq:R_branch}
\bar{\nu}_{R} = \Delta E_i^{TOSH} + \frac{1}{2}\left(A_1 + C_1\right)\left(J + 1\right)\left(J + 2\right) + \frac{1}{2}\left(A_1 - C_1\right) E_{\tau'} - \frac{1}{2}\left(A_0 + C_0\right)J\left(J + 1\right) - \frac{1}{2}\left(A_0 - C_0\right)E_{\tau''}
\end{equation}

As shown by these equations, for the frequency parameter we use the calculated anharmonic TOSH frequencies, $\Delta E_i^{TOSH}$, which correspond to mode $i$. Frequencies calculated by any other means can be used instead if wanted. Additionally, in these equations we have A$_1$, B$_1$, C$_1$ which are the rotational constants for the upper vibrationally excited level whereas A$_0$, B$_0$, C$_0$ are the rotational constants for the ground state. 

We quickly notice the need to derive E$_{\tau}$ values for the upper and lower levels, shown as E$_{\tau'}$ and E$_{\tau''}$ respectively. These parameters describe the energies of the sub-levels, which are observed in the asymmetric rotor spectra as the usual equations to describe degeneracy (for other types of polyatomic molecules such as symmetric and spherical) are no longer an accurate representation of the energy levels \citep[][]{12CoOhXX}{}{}. The sub-level values are calculated using an asymmetry parameter, which depends on $A$, $B$ and $C$, therefore meaning the above equations, \ref{eq:P_branch}, \ref{eq:Q_branch} and \ref{eq:R_branch} require all rotational constants. A typical energies diagram for an asymmetric molecule is shown in \citet{45HeXXXX} figure 17, which also shows its relation to the energy levels of a symmetric molecule.

Deriving the various sub-level energy values requires several steps: forming an energy matrix $\textbf{E} \left(\kappa\right)$, diagonalizing and then organising the energies, E$_{\tau}$, from lowest to highest to correspond correctly with the sub-levels. Also consideration needs to be taken as to what axis orientation is used for the system, as this will change the order of constants used within the formation of the energy matrix. In the case of ozone, which is a near-prolate asymmetric top, we have selected the I$^R$ format. I$^R$, specifically the I typing is mostly used when an asymmetric molecule is prolate-like in nature, and this parameter changes some of the values used in the calculation of the energy matrix. We will provide more implementation details in a forthcoming paper which describes the Pandora code, but the classic papers of \citet[][]{43KiHaCr}, \citet{45HeXXXX} and \citet{48GoXXXX} contain an extensive description of the theory used.

Additional selection rules need to be applied depending on the symmetry of the sub-level and the molecule's overall nuclear spin. We used the rules specified by \citet{97FlBaXX}, which depend on the type of band a vibrational mode possesses, the parity of the sub-levels, whether the isotopic variant is symmetric or asymmetric and which oxygen isotopes are present. The excellent book by \citet{63AlCrXX} goes into detail about the selection rules for the sub-levels as a function of band type and has also been used to confirm the required selection rules for ozone. The vibration bands typing for the modes have been manually set within our work.
 
Once the possible transitions and their frequencies have been calculated for each mode, we now must produce a means to approximately model the intensities. Following \citet{22CrBePi}, we used a Boltzmann distribution with some modifications. The main modification is the adaptation of the energy component in the equation to now reflect the addition of E$_{\tau}$ and  multiple rotational constants rather than just the solitary rotational constant $B$. The full equation we use is:
\begin{equation}
\label{eq:boltz_distribution_expanded}
    \frac{n_J}{\sum n_J} =\frac{(2J+1)}{f}e^{\left(\frac{\frac{1}{2}\left(A + C\right)J\left(J + 1\right) + \frac{1}{2}\left(A - C\right)E_\tau}{kT}\right)}
\end{equation}

Where $n_J$ is the occupation of rotational level $J$ and $f$ represents the rotational partition function. Once all of these components are calculated, we combine them to plot the spectrum. Given the usage of a Boltzmann distribution for our intensities, we have to make sure that the comparison spectrum is converted into the same arbitrary relative units rather than the typical atmospheric units that most databases use.

\subsection{Computational Details}
\label{subsection:Computational Details}
Two different potential energy surfaces (PES) for ozone are used within this work, one obtained from \citet{18PoZoMi} and the other from \citet{13DaLoLi}. The surfaces have been labeled with the abbreviations 18PoZo and 13DaLo, respectively, throughout this paper. We show that using two separate PES help illustrate how the anharmonic corrections are heavily reliant on the accuracy of the PES. Indeed, \citet[][]{23SpSeMa}{}{} showed that the vibrational frequencies of ozone are difficult to model using perturbation-based methods. They suggest that this could originate from the subtle electronic structure of the molecule. It is also worth noting the literature values from \citet{20GaTeKe} that we use as comparison in Section \ref{subsection:comparisons_constants}, was obtained by the authors using the \citet{13DaLoLi} potential. 

The ozone equilibrium geometry was taken from the experimental data in \citet{90LeScXX} and \citet{06HiKiCh}. These provide the bond angle, $\theta$, as 116.8$^{\circ}$ and the oxygen to oxygen bond length as 1.272~\AA. The Born-Oppenheimer approximation leads us to the conclusion that the equilibrium geometry and the molecular potential function is the same for all isotopologues \citep[][]{76HeStXX}, so the same geometry from \citet{06HiKiCh} is used for all of the isotopic variants of ozone within this paper.

The diagonal elements of the Cartesian Hessian matrix in eq. \ref{eq:hessian_CART_expanded} are computed using:
\begin{equation}
\label{eq:eta_ii}
    \eta_{ii} = \frac{1}{12h^2}\left(-E_{\left(-2,0\right)} + 16E_{\left(-1,0\right)} - 30E_{\left(0,0\right)} + 16E_{\left(1,0\right)} - E_{\left(2,0\right)}\right)
\end{equation}

Whereas non-diagonal elements are described as:
\begin{equation}
\label{eq:eta_ij}
    \eta_{ij} = \frac{1}{4h^2}\left(
    E_{\left(-1,-1\right)} - E_{\left(-1,1\right)} - E_{\left(1,-1\right)} + E_{\left(1,1\right)}\right)
\end{equation}

The constant $h$ represents the step taken to displace the specific atoms, in our work we arbitrarily selected a value of 0.005~Bohrs (atomic units), from the stable range of values reported in \citet{00AsRuTa}. The $E_{(a,b)}$ describes the energy calculated at a nuclear configuration displaced by $ah + bh$ from equilibrium.

The cubic and quartic constants required to calculate the TOSH anharmonically corrected band origin (eq. \ref{eq:Delta_E_TOSH}) are obtained through a method similar to the second order derivatives for the Cartesian Hessian. 

The cubic diagonal elements are computed via the following equation:
\begin{equation}
\label{eq:eta_iii}
    \eta_{iii} = \frac{1}{2h_i^3} \left( -E_{\left(-2,0\right)} + 2E_{\left(-1,0\right)} - 2E_{\left(1,0\right)} + E_{\left(2,0\right)} \right)
\end{equation}

The cubic non-diagonal elements are expressed as:
\begin{equation}
\label{eq:eta_iij}
    \eta_{iij} = \frac{1}{2h_i^2h_j} \left( - E_{\left(-1,-1\right)} + E_{\left(-1,1\right)} + 2E_{\left(0,-1\right)} - 2E_{\left(0,1\right)} - E_{\left(1,-1\right)} + E_{\left(1,1\right)} \right)
\end{equation}

The quartic diagonal elements are calculated using:
\begin{equation}
\label{eq:eta_iiii}
    \eta_{iiii} = \frac{1}{h_i^4} \left(E_{\left(-2,0\right)} - 4E_{\left(-1,0\right)} + 6E_{\left(0,0\right)} - 4E_{\left(1,0\right)} + E_{\left(2,0\right)} \right)
\end{equation}

Finally, the quartic non-diagonal elements are as follows:
\begin{equation}
\label{eq:eta_iijj}
    \eta_{iijj} = \frac{1}{h_i^2h_j^2} ( E_{\left(-1,-1\right)} - 2E_{\left(-1,0\right)} + E_{\left(-1,1\right)} - 2E_{\left(0,-1\right)} + 4E_{\left(0,0\right)} - 2E_{\left(0,1\right)} + E_{\left(1,-1\right)} - 2E_{\left(1,0\right)} + E_{\left(1,1\right)} )
\end{equation}

Where $h_i = h / \sqrt\omega_i$ and $h_j = h / \sqrt\omega_j$ are the adaptive steps along the $Q_i$ and $Q_j$ normal coordinate \citep[][]{19ErMaFe}. The inclusion of adaptive steps, which depend on the normal coordinates, is a slight modification from the theory of \citet{08LiGiGI}. For cubic and quantic constant calculations, the $E_{\left(a,b\right)}$ within \crefrange{eq:eta_iii}{eq:eta_iijj} represent the energy calculated at a nuclear configuration displaced by $ah_iQ_i + bh_jQ_j$ from equilibrium.

A tabular representation of the cubic matrix for a hypothetical molecule with 3 modes is shown in table \ref{tab:cubic_store}, to highlight how we store the constants as a compressed two-dimensional matrix. A representation of the quartic matrix for the same system in a similar compressed representation is shown in table \ref{tab:quartic_store}.

\begin{table}[ht]
\begin{minipage}{.5\linewidth}

\centering

\begin{tabular}{|cc|ccc|}
\hline
\multicolumn{2}{|c|}{\multirow{2}{*}{$\eta_{iij}$}} & \multicolumn{3}{c|}{j}                                    \\ \cline{3-5} 
\multicolumn{2}{|c|}{}                              & 0                        & 1                        & 2   \\ \hline
\multicolumn{1}{|c|}{\multirow{3}{*}{ii}}    & 00   & \multicolumn{1}{c|}{000} & 001                      & 002 \\ \cline{3-4}
\multicolumn{1}{|c|}{}                       & 11   & \multicolumn{1}{c|}{110} & \multicolumn{1}{c|}{111} & 112 \\ \cline{4-5} 
\multicolumn{1}{|c|}{}                       & 22   & 220                      & \multicolumn{1}{c|}{221} & 222 \\ \hline
\end{tabular}
\caption{Cubic constants matrix storage scheme.}
\label{tab:cubic_store}

\end{minipage}\hfill
\begin{minipage}{.5\linewidth}

\begin{tabular}{|cc|ccc|}
\hline
\multicolumn{2}{|c|}{\multirow{2}{*}{$\eta_{iijj}$}} & \multicolumn{3}{c|}{jj}                                      \\ \cline{3-5} 
\multicolumn{2}{|c|}{}                               & 00                        & 11                        & 22   \\ \hline
\multicolumn{1}{|c|}{\multirow{3}{*}{ii}}    & 00    & \multicolumn{1}{c|}{0000} & 0011                      & 0022 \\ \cline{3-4}
\multicolumn{1}{|c|}{}                       & 11    & \multicolumn{1}{c|}{1100} & \multicolumn{1}{c|}{1111} & 1122 \\ \cline{4-5} 
\multicolumn{1}{|c|}{}                       & 22    & 2200                      & \multicolumn{1}{c|}{2211} & 2222 \\ \hline
\end{tabular}
\caption{Quartic constants matrix storage scheme.}
\label{tab:quartic_store}

\end{minipage}
\end{table}

\section{Results and Discussion: Spectroscopic Constants}
\label{section:ResultsA}

Firstly in Section \ref{subsection:comparisons_constants}, we compare the spectroscopic constants from literature and those calculated by Epimetheus using two different potentials one from \citet{13DaLoLi} and one from \citet{18PoZoMi}. 

Following this, Section \ref{subsection:other_bandorigins} shows how our calculated band origins compare with the rest of the isotopic variants of ozone. This includes the ozone isotopologues 678, 786, 768, 677, 767, 788, 878, 778, 787. Only the band origins have been assessed for those as little data exist for rotational constants, presumably due to the low abundances of those isotopologues.
  
\subsection{Comparisons of band origins and rotational constants}
\label{subsection:comparisons_constants}

Table \ref{tab:constant_comparisons_666_and_668} displays the spectroscopic constants for ozone 666 ($^{16}$O$_3$) and 668 ($^{16}$O$^{16}$O$^{18}$O). Note that the table columns for each isotopic variant is organised as follows from left to right: the literature value (from either \citet{97FlBaXX} or \citet{20GaTeKe} typically) followed by the value from the \citet{18PoZoMi} potential and its difference to the literature values and finally the value from the \citet{13DaLoLi} potential and its difference to the literature values. The rows are organised as follow: we first cover the ground or ``000" state, starting with the $A$ rotational constant then the $B$ and the C. The same pattern is then repeated with the ``100", ``010" and ``001" states with the exception of the harmonic frequency, $\omega_i$, and the anharmonic frequency, $\nu_i$, values preceding the rotational constants. Where no data is present has been indicated by the $-$ character.

A key point to note here is that they are symmetric and asymmetric versions of ozone respectively and therefore those correspond to different point symmetry groups. This is a factor that will become more important further into the discussion.

\begin{table}[htb]
\centering
\begin{tabular}{|c|c||c|cc|cc||c|cc|cc|} 
\hline
\multirow{3}{*}{Mode} & \multirow{3}{*}{{Constant}} & \multicolumn{5}{c||}{666 ($^{16}$O$_3$) $\dagger$}                                           & \multicolumn{5}{c|}{668 ($^{16}$O$^{16}$O$^{18}$O) $\dagger$}                                           \\ 
\cline{3-12}
                      &                                     & Lit.   & \multicolumn{4}{c||}{Epimetheus}                           & Lit.   & \multicolumn{4}{c|}{Epimetheus}                           \\\cline{3-12}
                      &                                     & 97FlBa & 18PoZo & \multicolumn{1}{c}{$\Delta$Lit.} & 13DaLo & $\Delta$Lit. & 20GaTe & 18PoZo & \multicolumn{1}{l}{$\Delta$Lit.} & 13DaLo & $\Delta$Lit  \\ 
\hhline{|============|}
\multirow{3}{*}{000}  & $A$                                   & 3.554  & 3.557  & $-$0.003                       & 3.557  & $-$0.003   & 3.488  & 3.491  & +0.003                       & 3.491  & +0.003   \\
                      & $B$                                   & 0.445  & 0.449  & +0.004                       & 0.449  & +0.004   & 0.420  & 0.424  & +0.004                       & 0.424  & +0.004   \\
                      & $C$                                   & 0.395  & 0.399  & +0.004                       & 0.399  & +0.004   & 0.374  & 0.378  & +0.004                       & 0.378  & +0.004   \\ 
\hhline{|============|}
\multirow{5}{*}{100}  & $\omega_1$                               & 1134.9$^{\star}$      & 1134.9 & -                           & 1122.4 & $-$12.5       & -      & 1122   & -                           & 1112   & -       \\
                      & $\nu_1$                                  & 1103.1 & 1121.7 & +18.6                        & 1101.7 & $-$1.4     & 1090   & 1101   & +11                          & 1083.2 & $-$6.8     \\ \cline{2-12}
                      & $A$                                   & 3.557  & 3.511  & $-$0.046                       & 3.512  & $-$0.045   & 3.498  & 3.479  & $-$0.019                       & 3.311  & $-$0.187   \\
                      & $B$                                   & 0.443  & 0.442  & $-$0.001                       & 0.442  & $-$0.001   & 0.418  & 0.419  & +0.001                       & 0.423  & +0.005   \\
                      & $C$                                   & 0.393  & 0.393  & 0.0                          & 0.393 & 0.0      & 0.376  & 0.374  & $-$0.002             & 0.375  & $-$0.001   \\ 
\hhline{|============|}
\multirow{5}{*}{010}  & $\omega_2$                               & 716.0$^{\star}$     & 715.5  & $-$0.5                           & 708.7  & $-$7.3       & -      & 698.7  & -                           & 692.2  & -       \\
                      & $\nu_2$                                  & 700.9  & 702.4  & +1.5                         & 690.8  & $-$10.1    & 684.6  & 686.2  & +1.6                         & 675.7  & $-$8.9     \\ \cline{2-12}
                      & $A$                                   & 3.607  & 3.530  & $-$0.077                       & 3.530  & $-$0.077   & 3.549  & 3.490  & $-$0.059                       & 3.653  & +0.096   \\
                      & $B$                                   & 0.444  & 0.444  & 0.0                          & 0.443  & $-$0.001   & 0.419  & 0.419  & 0.0                          & 0.423  & +0.004   \\
                      & $C$                                   & 0.392  & 0.394  & +0.002                       & 0.394  & +0.002   & 0.372  & 0.374  & +0.002                       & 0.379  & +0.007   \\
\hhline{|============|}
\multirow{5}{*}{001}  & $\omega_3$                               & 1089.2$^{\star}$      & 1090.0 & +0.8                           & 1089.6 & +0.4       & -      & 1073   & -                           & 1071   & -       \\
                      & $\nu_3$                                  & 1042.1 & 1046.9 & +4.8                         & 1039.0 & $-$3.1     & 1028   & 1032   & +4                           & 1027   & +1       \\ \cline{2-12}
                      & $A$                                   & 3.501  & 3.557  & +0.056                       & 3.557  & +0.056   & 3.451  & 3.462  & +0.011                       & 3.442  & $-$0.009   \\
                      & $B$                                   & 0.441  & 0.449  & +0.008                       & 0.449  & +0.008   & 0.416  & 0.424  & +0.008                       & 0.424  & +0.008   \\
                      & $C$                                   & 0.391  & 0.399  & +0.008                       & 0.399  & +0.008   & 0.366  & 0.378  & +0.012                       & 0.378  & +0.012   \\ 
\hline
\end{tabular}
\caption{The constants for ozone 666 ($^{16}$O$_3$) and ozone 668 ($^{16}$O$^{16}$O$^{18}$O), all given in wave numbers (cm$^{-1}$). 97FlBa refers to \citet{97FlBaXX}, 20GaTe refers to \citet{20GaTeKe}, 18PoZo refers to \citet{18PoZoMi} and 13DaPh refers to \citet{13DaLoLi}. The $\dagger$ next to entries indicates that there are spectra available for comparison from HITRAN. The constant $\omega$ refers to the harmonic frequency of the mode whereas $\nu$ refers to the anharmonically corrected frequency. A $^{\star}$ next to value indicates the constant has been taken from \citet{74BaSeJo}.}
\label{tab:constant_comparisons_666_and_668}
\end{table}

First, focusing on the rotational constants for 666 (Table \ref{tab:constant_comparisons_666_and_668}), we see that there is little variation in the results obtained between the two potentials. Thus, we mainly highlight the comparisons to the literature, which pertain to both of the Epimetheus results. Overall, typically the obtained rotational constants are consistent with the literature well within 3\%. In particular, the $B$ and $C$ rotational constants are usually reproduced within a $\pm$0.002~cm$^{-1}$ difference to the literature. The $A$ rotational constant shows a difference from the literature ranging from 0.045~cm$^{-1}$ to 0.077~cm$^{-1}$. However, the $A$ rotational constant is an order of magnitude larger than $B$ and $C$, one would expect the difference from literature value for $A$ to also be at least a magnitude larger, if the accuracy is of a similar level.

If we now consider the constants for 668 (Table \ref{tab:constant_comparisons_666_and_668}), unlike with 666, by using different potentials we find a significant impact on the obtained rotational constants, with the 18PoZo potential typically providing more accurate results. For example, for rotational constant $A$ of mode 100, 18PoZo gives a difference of $-0.019$~cm$^{-1}$, whereas 13DaLo has a difference of $-0.187$~cm$^{-1}$. This is nearly an order of magnitude different and quite a large discrepancy from the literature \citep[][]{20GaTeKe}. Potentially this is due to the refinement of the potential, as it is five years younger than 13DaLo, and therefore may have utilised better methodologies in it's creation.

Since the harmonic frequencies, indicated by the entries for $\omega_i$ (rather than just the anharmonic frequencies, $\nu_i$), are also available in the literature, we can also use them to verify the accuracy of our method and/or the potential energy surface. Let us consider here the vibrational frequencies for 666. We obtain that by using 18PoZo Epimetheus reproduces the harmonic frequencies within a single wave number of the benchmark values from \citet{06HiKiCh}. This is not the case for 13DaLo: we can indeed reproduce the mode 001 harmonic value within a wave number, but are unable to replicate this accuracy for both modes 100 and 010, with differences of $-12.6$~cm$^{-1}$ and $-7.3$~cm$^{-1}$, respectively.

The comparison become more complicated if we consider the anharmonic frequencies, $\nu_i$. When using 18PoZo we no longer have a general trend of high accuracy, but instead a variation depending on the mode. For example, for mode 010 (which is the best result) we only have a discrepancy of $+1.5$~cm$^{-1}$. On the other hand, for mode 100 we obtain a difference of $+18.6$~cm$^{-1}$. Interestingly, by using 18PoZo for 666 all $\nu_i$ values have a positive difference from \citet{97FlBaXX}. This would suggest that our derived anharmonic corrections are smaller than they should be, since the harmonic values for that potential reproduce the literature values very closely. This is an issue which could be potentially reduced within Epimetheus if other methods for calculating the cubic and quartic constants are used.

With 13DaLo we instead obtain inverse results compared to 18PoZo: mode 100 is the best result with a difference of $-1.4$~cm$^{-1}$, while mode 010 is the worst with $-10.1$~cm$^{-1}$. In general, it appears that all the $\nu_i$ values for 13DaLo are over-corrected, potentially arising from the fact that the harmonic frequencies are mostly underestimated for this PES.

Finally, the vibrational frequencies for 668 exhibit similar correction patterns to 666, i.e., 18PoZo typically under-corrects the frequencies and leads to positive differences, whereas 13DaLo over-corrects causing negative differences. The main difference between 666 and 668 is the size of the corrections for the various modes. For example, for mode 100 666 has $\nu_{1}$ deviating by $+18.6$~cm$^{-1}$ (18PoZo) and $-1.4$~cm$^{-1}$ (13DaLo), whereas 668 deviates by $+11$~cm$^{-1}$ (18PoZo) and $-6.8$~cm$^{-1}$ (13DaLo), for the same mode. In contrast, the predictions for modes 010 and 001 are of similar quality for both 666 and 668.

\begin{table}[htb]
\centering
\begin{tabular}{|c|c||c|cc|cc||c|cc|cc|} 
\hline
\multirow{3}{*}{Mode} & \multirow{3}{*}{Constant} & \multicolumn{5}{c||}{686 ($^{16}$O$^{18}$O$^{16}$O) $\dagger$}                       & \multicolumn{5}{c|}{667 ($^{16}$O$^{16}$O$^{17}$O) $\dagger$}                       \\ 
\cline{3-12}
                      &                           & Lit.   & \multicolumn{4}{c||}{Epimetheus}       & Lit.   & \multicolumn{4}{c|}{Epimetheus}       \\ \cline{3-12}
                      &                           & 20GaTe & 18PoZo & $\Delta$Lit. & 13DaLo & $\Delta$Lit. & 97FlBa & 18PoZo & $\Delta$Lit. & 13DaLo & $\Delta$Lit  \\ 
\hhline{|============|}
\multirow{3}{*}{000}  & $A$                         & 3.290  & 3.293  & +0.003   & 3.293  & +0.003   & 3.519  & 3.522  & +0.003   & 3.522  & +0.003   \\
                      & $B$                         & 0.445  & 0.449  & +0.004   & 0.449  & +0.004   & 0.432 & 0.436 & +0.004   & 0.436  & +0.004   \\
                      & $C$                         & 0.391  & 0.395  & +0.004   & 0.395  & +0.004   & 0.384  & 0.388  & +0.004   & 0.388  & +0.004   \\ 
\hhline{|============|}
\multirow{5}{*}{100}  & $\omega_1$                     & -      & 1105   & -       & 1093   & -       & -      & 1127.3 & - & 1115.4 & -       \\
                      & $\nu_1$                        & 1074   & 1092   & +18      & 1073   & $-$1       & 1095.7 & 1111.4 & +15.7    & 1091.7 & $-$4.0     \\\cline{2-12}
                      & $A$                         & 3.299  & 3.256  & $-$0.043   & 3.256  & $-$0.043   & 3.522  & 3.452  & $-$0.070   & 3.346  & $-$0.176   \\
                      & $B$                         & 0.443  & 0.442  & $-$0.001   & 0.442  & $-$0.001   & 0.430  & 0.429  & $-$0.001   & 0.436  & +0.006   \\
                      & $C$                         & 0.393  & 0.389  & $-$0.004   & 0.389  & $-$0.004   & 0.382  & 0.381  & $-$0.001   & 0.386  & +0.004   \\ 
\hhline{|============|}
\multirow{5}{*}{010}  & $\omega_2$                     & -      & 707.1  & -       & 700.3  & -       & -      & 706.8  & -       & 700.1  & -       \\
                      & $\nu_2$                        & 696.3  & 694.5  & $-$1.8     & 683.2  & $-$13.1    & 692.4  & 694.0~ & +1.6     & 682.8  & $-$9.6     \\\cline{2-12}
                      & $A$                         & 3.351  & 3.276  & $-$0.075   & 3.275  & $-$0.076   & 3.570  & 3.479  & $-$0.091   & 3.672  & +0.102   \\
                      & $B$                         & 0.444  & 0.444  & 0.0      & 0.443  & $-$0.001      & 0.431  & 0.430  & $-$0.001   & 0.435  & +0.004   \\
                      & $C$                         & 0.389  & 0.391  & +0.002   & 0.391 & +0.002   & 0.382  & 0.383  & +0.001   & 0.389  & +0.007   \\
\hhline{|============|}
\multirow{5}{*}{001}  & $\omega_3$                     & -      & 1053   & -       & 1053   & -       & -      & 1082.2 & -       & 1081.3 & -       \\
                      & $\nu_3$                        & 1008   & 1013   & +5       & 1006   & $-$2       & 1035.4 & 1040.0 & +4.6     & 1033.4 & $-$6.6     \\\cline{2-12}
                      & $A$                         & 3.253  & 3.293  & +0.040   & 3.293  & +0.040   & 3.467  & 3.527  & +0.060   & 3.499  & +0.032   \\
                      & $B$                         & 0.441  & 0.449  & +0.008   & 0.449  & +0.008   & 0.428  & 0.435  & +0.007   & 0.436  & +0.008   \\
                      & $C$                         & 0.384  & 0.395  & +0.011   & 0.395  & +0.008   & 0.380  & 0.387  & +0.007   & 0.388 & +0.008   \\ 
\hline
\end{tabular}
\caption{The constants for ozone 686 ($^{16}$O$^{18}$O$^{16}$O) and ozone 667 ($^{16}$O$^{16}$O$^{17}$O), all given in wave numbers (cm$^{-1}$). 97FlBa refers to \citet{97FlBaXX}, 20GaTe refers to \citet{20GaTeKe}, 18PoZo refers to \citet{18PoZoMi} and 13DaPh refers to \citet{13DaLoLi}. The $\dagger$ next to entries indicates that there are spectra available for comparison from HITRAN. The constant $\omega$ refers to the harmonic frequency of the mode whereas $\nu$ refers to the anharmonically corrected frequency.}
\label{tab:constant_comparisons_686_and_667}
\end{table}

The spectroscopic constants for ozone 686 ($^{16}$O$^{18}$O$^{16}$O) and 667 ($^{16}$O$^{16}$O$^{17}$O) are shown in Table \ref{tab:constant_comparisons_686_and_667}. 
Similarly to 666 and 668 in Table \ref{tab:constant_comparisons_666_and_668}, they are respectively symmetric and asymmetric isotopologues of ozone, and therefore they correspond to different point symmetry groups. 

Focusing first on the rotational constants, we see a similar pattern as before. The rotational constants for the symmetric isotopologue 686 are mostly the same for the two potentials and are good approximations of the literature. This is not the case with the asymmetric 667, whose values vary more than 686. Overall, it appears that 18PoZo produces slightly better rotational constants for 667. For example, the rotational constants for mode 001 obtained with 18PoZo differ from the literature by $-0.091$~cm$^{-1}$, $-0.001$~cm$^{-1}$ and $+0.001$~cm$^{-1}$, for $A$, $B$ and $C$, whilst 13DaLo gives differences of $+0.102$~cm$^{-1}$, $+0.004$~cm$^{-1}$ and $+0.007$~cm$^{-1}$.

The 18PoZo PES for 686 produces anharmonic frequencies that deviate in a similar manner to those of 666. This is characterised mainly by the larger difference for mode 100 in the region of $18$~cm$^{-1}$, then an extremely accurate result for mode 010 of less than $2$~cm$^{-1}$ followed by a value roughly 5~cm$^{-1}$ off the literature for mode 001. When using 13DaLo PES for 686, the results are similar to 666: with modes 100 and 001 we obtain good results, less than a couple of wave numbers off, with a deterioration in accuracy for mode 010. 

\begin{table}[htb]
\centering
\begin{tabular}{|c|c||c|cc|cc||c|cc|cc|} 
\hline
\multirow{3}{*}{Mode} & \multirow{3}{*}{Constant} & \multicolumn{5}{c||}{676 ($^{16}$O$^{17}$O$^{16}$O) $\dagger$}                       & \multicolumn{5}{c|}{688 ($^{16}$O$^{18}$O$^{18}$O)}                        \\ 
\cline{3-12}
                      &                           & Lit.   & \multicolumn{4}{c||}{Epimetheus}       & Lit.   & \multicolumn{4}{c|}{Epimetheus}        \\ \cline{3-12}
                      &                           & 97FlBa & 18PoZo & $\Delta$Lit. & 13DaLo & $\Delta$Lit. & 97FlBa & 18PoZo  & $\Delta$Lit. & 13DaLo & $\Delta$Lit  \\ 
\hhline{|============|}
\multirow{3}{*}{000}  & $A$                         & 3.414  & 3.417  & +0.003   & 3.417  & +0.003   & 3.225  & 3.227   & +0.002   & 3.227  & +0.002   \\
                      & $B$                         & 0.445  & 0.449  & +0.004   & 0.449  & +0.004   & 0.420  & 0.423   & +0.003   & 0.423  & +0.003   \\
                      & $C$                         & 0.393  & 0.397  & +0.004   & 0.397  & +0.004   & 0.371  & 0.374   & +0.003   & 0.374  & +0.003   \\ 
\hhline{|============|}
\multirow{5}{*}{100}  & $\omega_1$                      & -      & 1118.9 & -       & 1106.8 & -       & -      & 1091.6  & -       & 1080.9 & -       \\
                      & $\nu_1$                        & 1087.8 & 1106.2 & +18.4    & 1086.6 & $-$1.2     & 1060.7 & 1072.2 & +11.5    & 1055.0 & $-$5.7     \\ 
\cline{2-12}
                      & $A$                         & 3.415 & 3.376  & $-$0.039   & 3.377  & $-$0.038   & -      & 3.159   & -        & 3.065  & -        \\
                      & $B$                         & 0.443  & 0.442  & $-$0.001   & 0.442  & $-$0.001   & -      & 0.416   & -        & 0.423  & -        \\
                      & $C$                         & 0.391  & 0.391  & 0.0      & 0.391  & 0.0      & -      & 0.368   & -        & 0.372  & -        \\
\hhline{|============|}
\multirow{5}{*}{010}  & $\omega_2$                      & -      & 711.2  & -       & 704.4  & -       & -      & 690.9   & -       & 684.3  & -       \\
                      & $\nu_2$                        & 697.1 & 698.4  & +1.3     & 686.9  & $-$10.2    & 677.5  & 678.7   & +1.2     & 668.4  & $-$9.1     \\ 
\cline{2-12}
                      & $A$                         & 3.466  & 3.395  & $-$0.071   & 3.395  & $-$0.071   & 3.273  & 3.190   & $-$0.083   & 3.357  & +0.084   \\
                      & $B$                         & 0.444  & 0.444  & 0.0      & 0.444  & 0.0      & 0.419  & 0.418   & $-$0.001   & 0.423  & +0.004   \\
                      & $C$                         & 0.391  & 0.393  & +0.002   & 0.392  & +0.001   & 0.369  & 0.370   & +0.001   & 0.376  & +0.007   \\
\hhline{|============|}
\multirow{5}{*}{001}  & $\omega_3$                      & -      & 1070.8 & -       & 1070.4 & -       & -      & 1036.3  & -       & 1034.9 & -       \\
                      & $\nu_3$                        & 1024.4 & 1029.1 & +4.7     & 1021.6 & $-$2.8     & 993.9  & 998.2   & +4.3     & 993.1  & $-$0.8     \\ 
\cline{2-12}
                      & $A$                         & 3.364  & 3.417  & +0.053   & 3.417  & +0.053   & -      & 3.230   & -        & 3.189  & -        \\
                      & $B$                         & 0.441  & 0.449  & +0.008   & 0.449  & +0.008   & -      & 0.422   & -        & 0.424  & -        \\
                      & $C$                         & 0.390  & 0.397  & +0.007   & 0.397  & +0.007   & -      & 0.373   & -        & 0.374  & -        \\ 
\hline
\end{tabular}
\caption{The constants for ozone 676 ($^{16}$O$^{17}$O$^{16}$O) and ozone 688 ($^{16}$O$^{18}$O$^{18}$O), all given in wave numbers (cm$^{-1}$). 97FlBa refers to \citet{97FlBaXX}, 18PoZo refers to \citet{18PoZoMi} and 13DaPh refers to \citet{13DaLoLi}. The $\dagger$ next to entries indicates that there are spectra available for comparison from HITRAN. The constant $\omega$ is the harmonic frequency of the mode whereas $\nu$ is the anharmonically corrected frequency. The column title "Lit." naturally indicates the literature source of the target values.}
\label{tab:constant_comparisons_676_and_688}
\end{table}

Table \ref{tab:constant_comparisons_676_and_688} shows the spectroscopic constants for the ozone isotopologues 676 ($^{16}$O$^{17}$O$^{16}$O) and 668 ($^{16}$O$^{18}$O$^{18}$O). For 688, modes 100 and 001 have no reported data for their rotational constants. Similarly to the previous Tables \ref{tab:constant_comparisons_666_and_668}-\ref{tab:constant_comparisons_686_and_667}, we have a symmetric specie (676) and an asymmetric specie (688) of ozone. 
The results for 676 and 688 are very similar to what has been shown before. The only significant novelty now is that for 676 all fundamental frequencies are larger if 18PoZo is used, or smaller in case 13DaLo is implemented. 

\begin{table}[htb]
\centering
\begin{tabular}{|c|c||c|cc|cc||c|cc|cc|} 
\hline
\multirow{3}{*}{Mode} & \multirow{3}{*}{Constant} & \multicolumn{5}{c||}{868 ($^{18}$O$^{16}$O$^{18}$O)$\ddagger$}                       & \multicolumn{5}{c|}{888 ($^{18}$O$_3$)$\ddagger$}                       \\ 
\cline{3-12}
                      &                           & Lit.   & \multicolumn{4}{c||}{Epimetheus}       & Lit.   & \multicolumn{4}{c|}{Epimetheus}       \\ \cline{3-12}
                      &                           & 20GaTe & 18PoZo & $\Delta$Lit. & 13DaLo & $\Delta$Lit. & 97FlBa & 18PoZo & $\Delta$Lit. & 13DaLo & $\Delta$Lit  \\ 
\hhline{|============|}
\multirow{3}{*}{000}  & $A$                         & 3.422 & 3.427  & +0.005   & 3.427  & +0.005   & 3.158  & 3.161  & +0.003   & 3.161  & +0.003   \\
                      & $B$                         & 0.396  & 0.399  & +0.003   & 0.399  & +0.003   & 0.396  & 0.399  & +0.003   & 0.399  & +0.003   \\
                      & $C$                         & 0.354  & 0.357  & +0.003   & 0.357  & +0.003   & 0.351  & 0.354  & +0.003   & 0.354  & +0.003   \\ 
\hhline{|============|}
\multirow{5}{*}{100}  & $\omega_1$                     & -      & 1102   & -       & 1089   & -       & 1070.0$^{\star}$     & 1069.8 & -0.2       & 1058.1 & -11.9       \\
                      & $\nu_1$                        & 1072   & 1089   & +17      & 1070   & $-$2       & 1041.6 & 1058.1 & +16.5    & 1039.6 & $-$2       \\ 
\cline{2-12}
                      & $A$                         & 3.438 & 3.376  & $-$0.062   & 3.377  & $-$0.061   & 3.161  & 3.122  & $-$0.039   & 3.123  & $-$0.038   \\
                      & $B$                         & 0.394  & 0.394  & 0.0      & 0.393  & $-$0.001   & 0.394  & 0.393  & $-$0.001   & 0.393  & $-$0.001   \\
                      & $C$                         & 0.357  & 0.352  & $-$0.005   & 0.352  & $-$0.005   & 0.349  & 0.349  & 0.0      & 0.349  & 0.0      \\ 
\hhline{|============|}
\multirow{5}{*}{010}  & $\omega_2$                     & -      & 681.8  & -       & 675.4  & -       & 674.8$^{\star}$      & 674.5  & $-$0.3       & 668.1  & $-$6.7       \\
                      & $\nu_2$                        & 668.1 & 669.8  & +1.7     & 658.4  & $-$9.7     & 661.5  & 662.9  & +1.2     & 651.8  & $-$11.1    \\ 
\cline{2-12}
                      & $A$                         & 3.479  & 3.391  & $-$0.088   & 3.391  & $-$0.088   & 3.203  & 3.138  & $-$0.065   & 3.138  & $-$0.065   \\
                      & $B$                         & 0.395  & 0.394  & $-$0.001   & 0.394  & $-$0.001   & 0.395  & 0.395  & 0.0      & 0.394  & $-$0.001   \\
                      & $C$                         & 0.352  & 0.353  & +0.001   & 0.353  & +0.001   & 0.349  & 0.351  & +0.002   & 0.350  & +0.001   \\
\hhline{|============|}
\multirow{5}{*}{001}  & $\omega_3$                     & -      & 1065   & -       & 1065   & -       & 1026.5$^{\star}$      & 1027.5 & +1.0       & 1027.2 & +0.7       \\
                      & $\nu_3$                        & 1019   & 1024   & +5       & 1016   & $-$3       & 984.8  & 989.1  & +4.3     & 982.3  & $-$2.5     \\ 
\cline{2-12}
                      & $A$                         & 3.381  & 3.425  & +0.044   & 3.425  & +0.044   & 3.114  & 3.161  & +0.047   & 3.161  & +0.047   \\
                      & $B$                         & 0.392  & 0.399  & +0.007   & 0.399  & +0.007   & 0.393  & 0.399  & +0.006   & 0.399  & +0.006   \\
                      & $C$                         & 0.345  & 0.357  & +0.012   & 0.357  & +0.012   & 0.348  & 0.354  & +0.006   & 0.354  & +0.006   \\ 
\hline
\end{tabular}
\caption{The constants for ozone 868 ($^{18}$O$^{16}$O$^{18}$O) and ozone 888 ($^{18}$O$_3$), all given in wave numbers (cm$^{-1}$). 97FlBa refers to \citet{97FlBaXX}, 20GaTe refers to \citet{20GaTeKe}, 18PoZo refers to \citet{18PoZoMi} and 13DaPh refers to \citet{13DaLoLi}. The $\dagger$ next to entries indicates that there are spectra available for comparison from HITRAN. The constant $\omega$ is the harmonic frequency of the mode whereas $\nu$ is the anharmonically corrected frequency. The column title "Lit." naturally indicates the literature source of the target values. The $\ddagger$ indicates that comparison data from S\&MPO \citet{14BaMiBa} was available.  A $^{\star}$ next to a value indicates the constant has been taken from \citet{74BaSeJo}.}
\label{tab:constant_comparisons_868_and_888}
\end{table}

Table \ref{tab:constant_comparisons_868_and_888}, shows the constants for 868 ($^{18}$O$^{16}$O$^{18}$O) and 888 ($^{18}$O$_3$). For 888, we have the harmonic literature values from \citet{74BaSeJo}, allowing us a further insight to the abilities of the anharmonic corrections of the method. In particular, we 
show again an excellent agreement for harmonic frequencies when using the 18PoZo PES, while the results using the 13DaLo PES are not as good. Notice that Table \ref{tab:constant_comparisons_868_and_888} only contains symmetric isotopologues and this is reflected in the results we obtain. They match the patterns and general trends of the previous symmetric isotopologues shown by 676 (Table \ref{tab:constant_comparisons_676_and_688}), 686 (Table \ref{tab:constant_comparisons_686_and_667}) and 666 (Table \ref{tab:constant_comparisons_666_and_668}).

\begin{table}[htb]
\centering
\begin{tabular}{|c|c||c|cc|cc|} 
\hline
\multirow{3}{*}{Mode} & \multirow{3}{*}{Constant} & \multicolumn{5}{c|}{777 ($^{17}$O$_3$)}                        \\ 
\cline{3-7}
                      &                           & Lit.   & \multicolumn{4}{c|}{Epimetheus}        \\\cline{3-7}
                      &                           & 97FlBa & 18PoZo & $\Delta$Lit. & 13DaLo & $\Delta$Lit.  \\ 
\hhline{|=======|}
\multirow{3}{*}{000}  & $A$                         & 3.344 & 3.347  & +0.003   & 3.347  & +0.003    \\
                      & $B$                         & 0.419  & 0.423  & +0.004   & 0.423  & +0.004    \\
                      & $C$                         & 0.372  & 0.375  & +0.003   & 0.375  & +0.003    \\ 
\hhline{|=======|}
\multirow{5}{*}{100}  & $\omega_1$                     & -      & 1100.8 & -       & 1088.8 & -        \\
                      & $\nu_1$                        & 1070.9 & 1088.5 & +17.6    & 1069.3 & $-$1.6      \\ 
\cline{2-7}
                      & $A$                         & 3.347  & 3.304  & $-$0.043   & 3.306  & $-$0.041    \\
                      & $B$                         & 0.417  & 0.416  & $-$0.001   & 0.416  & $-$0.001    \\
                      & $C$                         & 0.369  & 0.370  & +0.001   & 0.370  & +0.001    \\ 
\hhline{|=======|}
\multirow{5}{*}{010}  & $\omega_2$                     & -      & 694.0  & -       & 687.5  & -        \\
                      & $\nu_2$                        & -      & 681.7  & -        & 670.4  & -         \\ 
\cline{2-7}
                      & $A$                         & -      & 3.322  & -        & 3.322  & -         \\
                      & $B$                         & -      & 0.418  & -        & 0.417  & -         \\
                      & $C$                         & -      & 0.371  & -        & 0.371  & -         \\
\hhline{|=======|}
\multirow{5}{*}{001}  & $\omega_3$                     & -      & 1057.3 & -       & 1056.9 & -        \\
                      & $\nu_3$                        & 1012.2 & 1016.7 & +4.5     & 1009.4 & $-$2.8      \\ 
\cline{2-7}
                      & $A$                         & 3.296  & 3.347  & +0.051   & 3.347  & +0.051    \\
                      & $B$                         & 0.415  & 0.423  & +0.008   & 0.423  & +0.008    \\
                      & $C$                         & 0.368  & 0.375  & +0.007   & 0.375  & +0.007    \\ 
\hline
\end{tabular}
\caption{The constants for ozone 777 ($^{17}$O$_3$), all given in wave numbers (cm$^{-1}$). 97FlBa refers to \citet{97FlBaXX}, 18PoZo refers to \citet{18PoZoMi} and 13DaPh refers to \citet{13DaLoLi}. The constant $\omega$ is the harmonic frequency of the mode whereas $\nu$ is the anharmonically corrected frequency. The column title "Lit." naturally indicates the literature source of the target values.}
\label{tab:constant_comparisons_777}
\end{table}

Finally, in Table \ref{tab:constant_comparisons_777} we compare the results for ozone 777 ($^{17}$O$_3$). No data, frequencies or rotational constants are available for the 010 mode of 777. We also have no harmonic frequencies, which means we can not evaluate how accurate the initial Hessian calculation are. Based on the data available for all other modes and the harmonic rotational constants, 
the results in Table \ref{tab:constant_comparisons_777} are consistent with the previous analysis.
As 777 is a symmetric variant, there is little difference in the rotational constants when using the two potentials. The frequency for 100 performs as expected, following the pattern seen for the other symmetric isotopologues: a discrepancy nearing $+20$~cm$^{-1}$ is obtained for 18PoZo, and a result of only a couple of wave numbers off for 13DaLo. The asymmetric stretch mode 001 results are what we also expect as they are not too dissimilar, being within 5~cm$^{-1}$ for both (but over for 18PoZo and under for 13DaLo).

\subsection{Other Isotopic Variants Band Origins}
\label{subsection:other_bandorigins}

\begin{table}[htb]
\centering
\begin{tabular}{|c|c||c||cc|cc|} 
\hline
\multirow{2}{*}{Isotopomer} & \multirow{2}{*}{Mode} & Literature      & \multicolumn{4}{c|}{Epimetheus}          \\ 
\cline{3-7}
                            &                       & 01HaMa & 18PoZo & $\Delta$Lit. & 13DaLo & $\Delta$Lit.  \\ 
\hhline{|=======|}
\multirow{3}{*}{678 ($^{16}$O$^{17}$O$^{18}$O)}        & 100                    & 1071.9   & 1085.8 & +13.9     & 1068.3 & $-$3.6       \\
                            & 010                   & 681.3    & 682.4  & +1.1      & 672.0  & $-$9.3       \\
                            & 001                   & 1012.9   & 1014.5 & +1.6      & 1009.3 & $-$3.6       \\ 
\hline
\multirow{3}{*}{786 ($^{17}$O$^{18}$O$^{16}$O)}        & 100                  & 1065.4   & 1082.1 & +16.7     & 1063.4 & $-$2.0       \\
                            & 010                   & 685.2    & 686.3  & +1.1      & 675.3  & $-$9.9       \\
                            & 001                   & 1001.9   & 1005.9 & +4.0      & 999.6  & $-$2.3       \\ 
\hline
\multirow{3}{*}{768 ($^{17}$O$^{16}$O$^{18}$O)}        & 100                   & 1079.8   & 1095.2 & +15.4     & 1075.7 & $-$4.1       \\
                            & 010                   & 676.1    & 677.7  & +1.6      & 666.6  & $-$9.5       \\
                            & 001                   & 1025.1   & 1028.7 & +3.6      & 1022.3 & $-$2.8       \\ 
\hline
\multirow{3}{*}{677 ($^{16}$O$^{17}$O$^{17}$O)}        & 100                    & 1079.6   & 1095.8 & +16.2     & 1076.7 & $-$2.9       \\
                            & 010                   & 688.9    & 690.1  & +1.2      & 679.0  & $-$9.9       \\
                            & 001                   & 1018.6   & 1022.1 & +3.5      & 1015.6 & $-$3.0       \\ 
\hline
\multirow{3}{*}{767 ($^{17}$O$^{16}$O$^{17}$O)}        & 100                    & 1087.2   & 1104.5 & +17.3     & 1084.8 & $-$2.4       \\
                            & 010                   & 684.0    & 685.5  & +1.5      & 674.0  & $-$10.0      \\
                            & 001                   & 1030.7   & 1034.7 & +4.0      & 1027.1 & $-$3.6       \\ 
\hline
\multirow{3}{*}{788 ($^{17}$O$^{18}$O$^{18}$O)}        & 100                    & 1048.7   & 1065.0 & +16.3     & 1046.5 & $-$2.2       \\
                            & 010                   & 669.0    & 670.5  & +1.5      & 659.6  & $-$9.4       \\
                            & 001                   & 989.5    & 994.1  & +4.6      & 988.0  & $-$1.5       \\ 
\hline
\multirow{3}{*}{878 ($^{18}$O$^{17}$O$^{18}$O)}        & 100                    & 1055.6   & 1072.8 & +17.2     & 1053.8 & $-$1.6       \\
                            & 010                   & 664.5    & 666.3  & +1.8      & 655.1  & $-$9.4       \\
                            & 001                   & 1000.6   & 1005.6 & +5.0      & 998.4  & $-$2.2       \\ 
\hline
\multirow{3}{*}{778 ($^{17}$O$^{17}$O$^{18}$O)}        & 100                    & 1063.3   & 1079.2 & +15.9     & 1060.2 & $-$3.1       \\
                            & 010                   & 672.5    & 674.0  & +1.5      & 663.1  & $-$9.4       \\
                            & 001                    & 1006.4   & 1010.5 & +4.1      & 1004.2 & $-$2.2       \\ 
\hline
\multirow{3}{*}{787 ($^{17}$O$^{18}$O$^{17}$O)}        & 100                    & 1056.5   & 1074.2 & +17.7     & 1055.5 & $-$1.0       \\
                            & 010                  & 676.7    & 678.1  & +1.4      & 666.9  & $-$9.8       \\
                            & 001                   & 995.4    & 1000.4 & +5.0      & 993.5  & $-$1.9       \\
\hline
\end{tabular}
\caption{The band origins for the remaining isotopologues of ozone. These are organised in order of decreasing abundance and all results have been presented in units of wave numbers (cm$^{-1}$). 01HaMa refers to \citet{01HaMaXX}, 18PoZo to \citet{18PoZoMi} and 13DaLo to \citet{13DaLoLi}.}
\label{tab:otherisoband}
\end{table}

In this section our results are compared with \citet{01HaMaXX}. Table \ref{tab:otherisoband} shows the remaining species in descending order of abundance. For these isotopomers all the data for each mode band origin is available. 

In general, the accuracy for each mode differs depending on the potential used, with reoccurring patterns similar to what we have seen in Section \ref{subsection:comparisons_constants}. However, neither 18PoZo or 13DaLo potentials present results that are substantially inaccurate, apart from mode 100 ($\nu_1$) for 18PoZo. Even this mode with the 18PoZo potential however always lies below 20~cm$^{-1}$ difference, with the asymmetric variants typically having a difference of less than 15~cm$^{-1}$. 

Concerning the $\nu_2$ mode, we see for 18PoZo the differences range from 1.1~cm$^{-1}$ to 1.8~cm$^{-1}$. Conversely, 13DaLo provides results that range 9.3~cm$^{-1}$ to 10.0~cm$^{-1}$. Interestingly, both potentials provide results that have roughly a constant difference across all isotopologues, which suggests the code provides a  consistent correction of this particular mode. 

In general, the differences between the accurate literature data and our results for ozone is mostly driven by the nature of the anharmonic corrections, instead of by the specific potential implemented. Indeed, the approach we have chosen is a simplifications of the second-order vibrational perturbation theory (VPT2) and truncates the energy expansion equation (see TOSH eqs. 21-24 in \citet{08LiGiGI}) used to derive our shift parameter $\sigma$. More specifically, some of the cubic constants (see eq. \ref{eq:eta_iii}, \ref{eq:eta_iij} and table \ref{tab:cubic_store}) that are traditionally included in the VPT2 method are not calculated or used within the TOSH framework. We think likely that the missing constants could have a noticeable contribution to the correction and thus lead to the inaccuracies in our approach. This issue presumably will only affect molecules of high symmetry, the reasons for which are discussed later in this Section.

For example, the effect of neglecting some VPT2 constants is striking for the results obtained with the 18PoZo potential: the harmonic values for this potential accurately reproduce the experimental data, yet the anharmonic corrections computed are too small to fully account for the observed anharmonicity. Conversely, the 13DaLo potential leads to better corrected values because the harmonic frequencies are underestimated initially. The limited anharmonic correction computed with our truncated approach has therefore less of a frequency gap to correct for.  

\begin{table}
\begin{tabular}{|c|c||ccc||ccc|} 
\hline
\multicolumn{1}{|c|}{\multirow{2}{*}{Source} }  &  \multirow{2}{*}{Mode}   & \multicolumn{3}{c||}{666 ($^{16}$O$_3$)}  & \multicolumn{3}{c|}{888 ($^{18}$O$_3$)}  \\ 
\cline{3-8}
\multicolumn{1}{|c|}{}   &     & Harmonic & Anharmonic & $\Delta$   & Harm   & Anharmonic & $\Delta$  \\ 
\hline\hline
\multirow{3}{*}{Exp. (98FlBa)} & 100 & 1134.9$^{\star}$ & 1103.1 & $-$31.8 & 1070.0$^{\star}$ & 1041.6 & $-$28.4 \\
                       & 010 & 716.0$^{\star}$ & 700.9 & $-$15.1 & 674.8$^{\star}$ & 661.5 & $-$13.3 \\
                       & 001 & 1089.2$^{\star}$ & 1042.1 & $-$47.1 & 1026.5$^{\star}$ & 984.8 & $-$41.7 \\ 
\hline\hline
\multirow{3}{*}{18PoZo}& 100 & 1134.9   & 1121.7 & $-$13.2 & 1069.8 & 1058.1 & $-$11.7 \\
                       & 010 & 715.5    & 702.4  & $-$13.1 & 674.5  & 662.9  & $-$11.6 \\
                       & 001 & 1090.0   & 1046.9 & $-$43.1 & 1027.5 & 989.1  & $-$38.4 \\ 
\hline
\multirow{3}{*}{13DaLo} & 100 & 1121.4   & 1101.7 & $-$19.7 & 1058.1 & 1039.6 & $-$18.5 \\
                       & 010 & 708.7    & 690.8  & $-$17.9 & 668.1  & 651.8  & $-$16.3\\
                       & 001 & 1089.6   & 1039.0 & $-$50.6 & 1027.2 & 982.3  & $-$44.9  \\
\hline
\end{tabular}
\caption{For ozone 666 and 888, a comparison between the harmonic and anharmonic values from Epimetheus is shown for each mode and each potential. 18PoZo refers to \citet{18PoZoMi} and 13DaPh refers to \citet{13DaLoLi}. Exp.\ 98FlBa represents the difference between the harmonic and anharmonic value from literature by \citet{97FlBaXX}. A $^{\star}$ next to a value indicates the constant has been taken from \citet{74BaSeJo}.}
\label{tab:anharmonic_corrections_666_888}
\end{table}

Table \ref{tab:anharmonic_corrections_666_888} shows the frequencies, both harmonic and anharmonic, that are produced by either PES in Epimetheus and the experimental literature values. The main aim of this table is to explore how the two potential energy surfaces predict the actual anharmonic shift, and how it then compares to actual anharmonic shift that is observed.

For both potentials, the 100 mode correction is substantially underestimated. The simulations using 13DaLo manage to recover about two thirds of the correction, whereas with 18PoZo we only recover about half of what was expected. For the 010 mode, the magnitude of the correction is consistent with the literature. For example, for ozone 666 we would expect a correction of 15.1~cm$^{-1}$, 18PoZo gives a correction of 13.1 whilst 13DaLo gives a correction of 17.9, meaning both are only a couple of wave numbers off. We notice from this example, that 18PoZo is under-correcting and 13DaLo is over correcting, something which is also true for the 010 mode for 888; 18PoZo has a $-$1.7~cm$^{-1}$ difference to the target value and 13DaLo is then +3~cm$^{-1}$. Moving onto the asymmetric mode 001, for 666 once again both potentials give results within approximately 4 wave numbers. Like before, with mode 010, 18PoZo underestimates the correction whereas 13DaLo overestimates and the same is also true for ozone 888. The trend emerging here may not be the case for all types of isotopomers. We only have a small sample of anharmonic corrections to compare with, and only for the symmetric variants of ozone.

The mode 100, which is the designation given to the symmetric stretch, shows the largest correction. The corrections rely heavily upon the calculations of the cubic constants, which gives rise to the shift parameter $\sigma$, and to a lesser degree the quartic. In particular, for the cubic constants in eqs., \ref{eq:eta_iii} \& \ref{eq:eta_iij} the equations are symmetric in terms of the shift displacements. This numerical approach is well adapted to symmetric modes as the various energy contributions computed along those modes will not cancel out. However, displacements along the asymmetric mode (mode 001) can lead to near-perfect cancellation of all terms for symmetric isotopomers and potentially cancel completely its anharmonic contribution. Ironically, this contribution is what is required to correct the frequencies of the symmetric modes 100 and 010, thus leading to errors for their anharmonic frequencies. Meanwhile, the asymmetric mode is corrected by contributions coming from the symmetric modes that are correctly calculated. We can observe this in the results from Table \ref{tab:anharmonic_corrections_666_888}: for either potential (18PoZo or 13DaLo), the correction for the symmetric stretch 100 is lower than expected, as it is composed of the correction due to its interaction with the (symmetric) bending mode 010 and the asymmetric stretch mode 001, which we presume to be nearly zero. The 010 bending mode is corrected by nearly the same amount, since it shares its anharmonic correction with the symmetric stretch mode (100). Likely, its interaction with the asymmetric stretch (001) is also near zero for the reasons stated above. The asymmetric stretch (001) is instead fully corrected by its interactions with both symmetric modes.  

In the case of an asymmetric molecule, we postulate that more terms survive in the calculation of the cubic and quartic constant with this new intrinsic asymmetry. This would explain why the results for mode 100 in the case of asymmetric ozone variants are typically better, although not as good as we would expect if this is the sole reason for the poor approximations of the anharmonic corrections. In table \ref{tab:otherisoband}, we see that when an asymmetric variant is present, the difference between literature and our calculations is reduced down to a couple of wave numbers for the symmetric mode. For example, 787 has a difference of 17.7~cm$^{-1}$ and 778 has a difference of 15.9~cm$^{-1}$. It is difficult to get a more precise assessment of the sources of discrepancy, as the harmonic values are not available in the literature for all isotopomers. With this information, we could see whether even the zero-order terms in these equations are correct, and whether the magnitude of the correction is well approximated.

\section{Results and Discussion: Spectral Modelling}
\label{section:ResultsB}

Following up from the analysis in the previous section, here we discuss the impact of the different constants obtained from Epimetheus using the potentials 18PoZo and 13DaLo, on the spectrum produced by Pandora for ozone 666. In particular, we compare our results with high-quality spectra such as the 2020 release from the High Resolution Transmission molecular spectroscopic database (HITRAN 2020).

HITRAN spectra are available from \citet{22GoRoHa} for ozone 666, 668, 686, 667, 676. For isotopomers 868, 888 the comparison spectra are obtained from the S\&MPO-2020d database \citep{14BaMiBa}. Whereas the isotopomers 688 and 777 do not have spectra available for comparisons, and therefore have been omitted from the spectral modelling.

As a brief rundown on the approximate regions of the modes in the spectrum for the isotopologues of ozone; mode 100 is expected to lie within 1050-1150cm$^{-1}$, mode 010 in 600-800cm$^{-1}$ and mode 001 in 950-1100cm$^{-1}$.

\subsection{Ozone 666}
\label{subsection:Results_O3_666}
Ozone 666 is the main isotopologue and a symmetric variant belonging to the point symmetry group C$_{2v}$. As such, it exhibits bands of either A or B typing for each fundamental mode. Since it is composed entirely of the $^{16}$O isotope, spin effects will appear. This means only levels of certain parity will be populated, as explained in \citet{97FlBaXX}.

\begin{figure}[ht]
    \centering
    \includegraphics[width=\columnwidth]{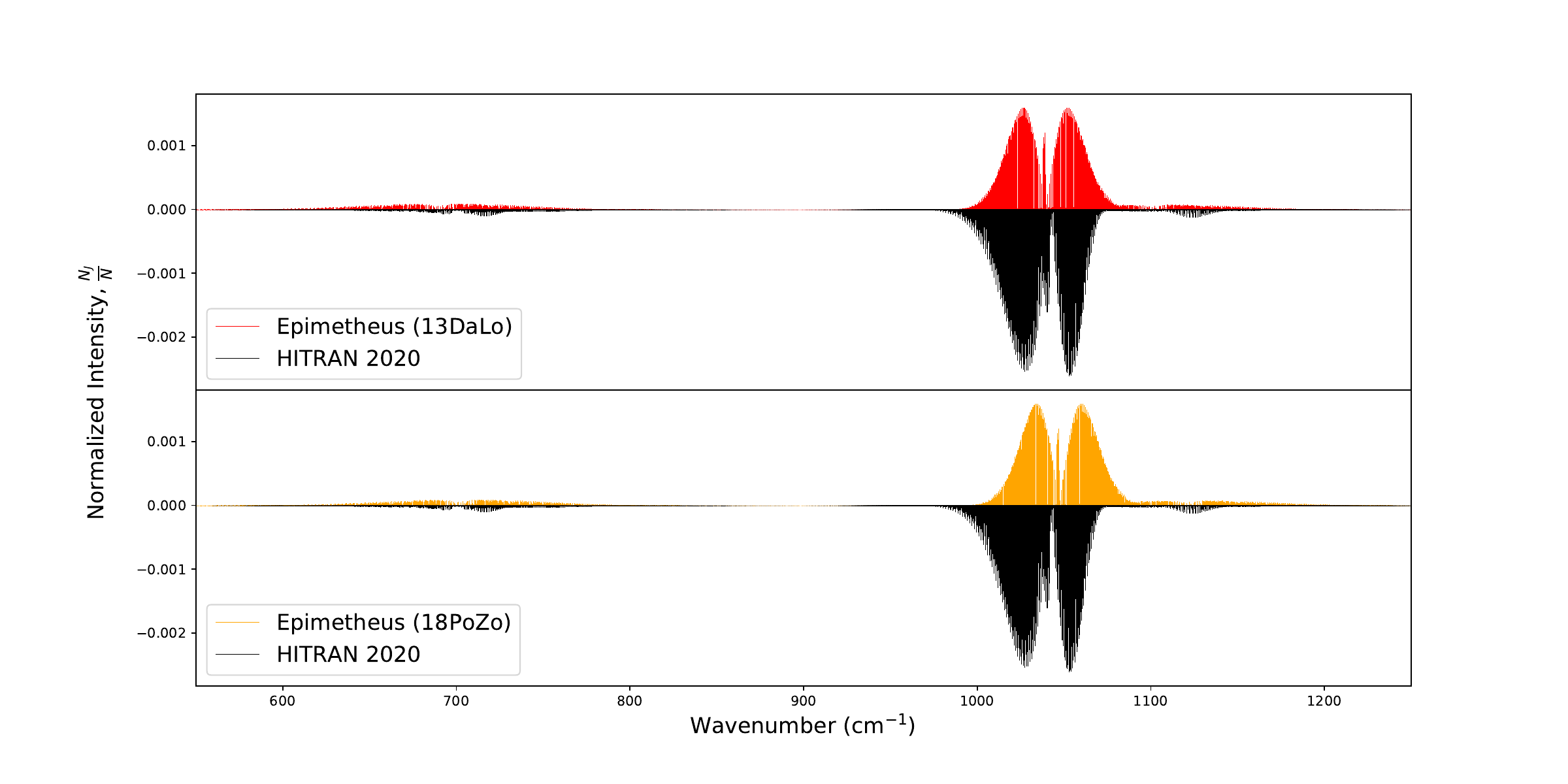}
    \caption{A comparison of the constants and ultimately the spectra obtained from both potentials from \cite{13DaLoLi} and \cite{18PoZoMi} for $^{16}$O$_3$.}
    \label{fig:EPI_O3_666_13DaLo_18PoZo_HITRAN_subplots_300K}
\end{figure}

\begin{figure}[htb]
    \centering
    \includegraphics[width=\columnwidth]{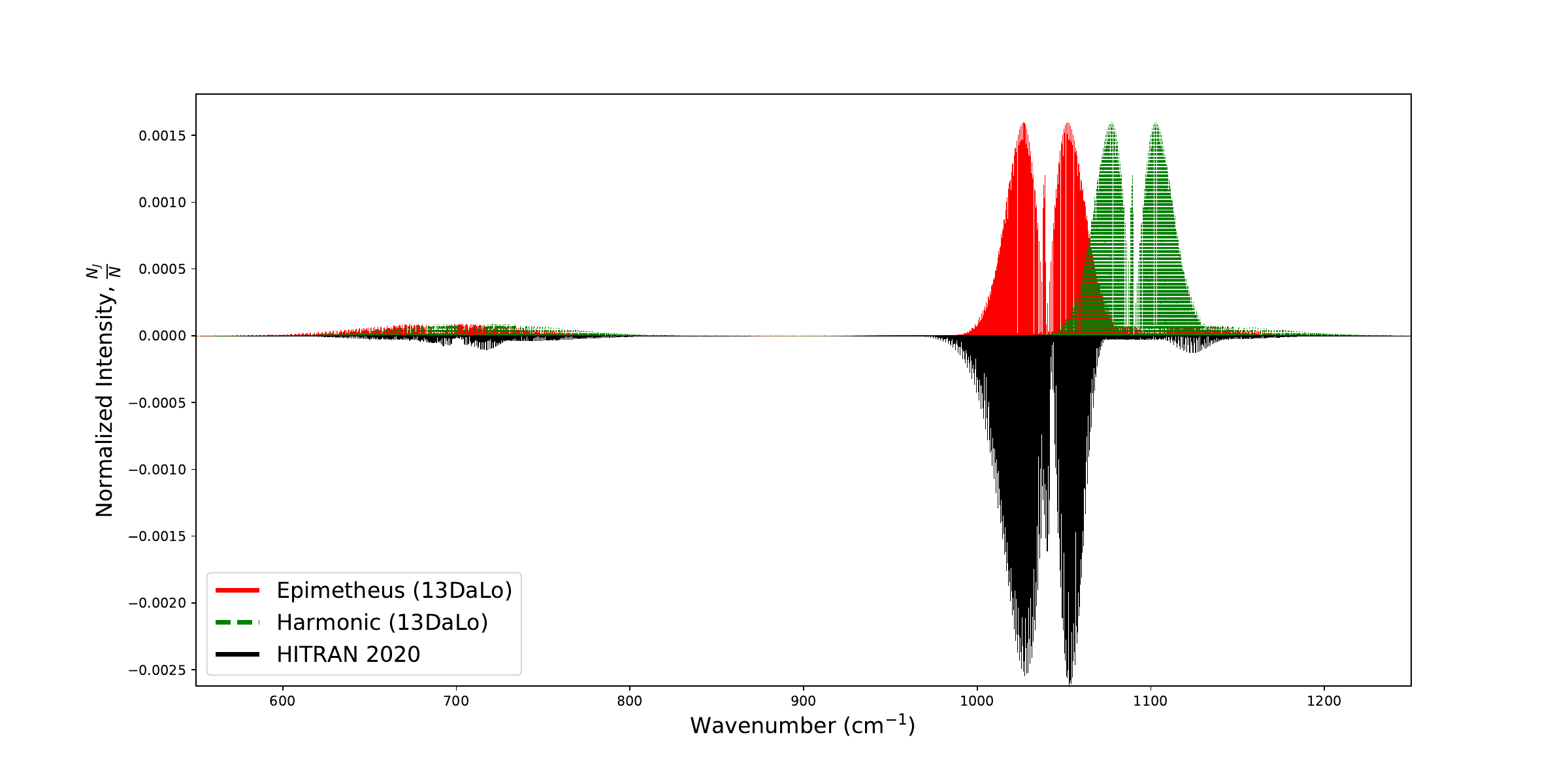}
    \caption{A comparison of the ozone 666 ($^{16}$O$_3$) spectrum from Epimetheus (13DaLo potential) in green. Data from the HITRAN 2020 release is in black and inverted for comparisons with our results. We have also included a harmonic spectrum also derived from the 13DaLo potential.}
    \label{fig:EPI_O3_666_13DaLo_HARMONIC_HITRAN2020_300K}
\end{figure}

\begin{figure}[htb]
    \centering
    \includegraphics[width=\columnwidth]{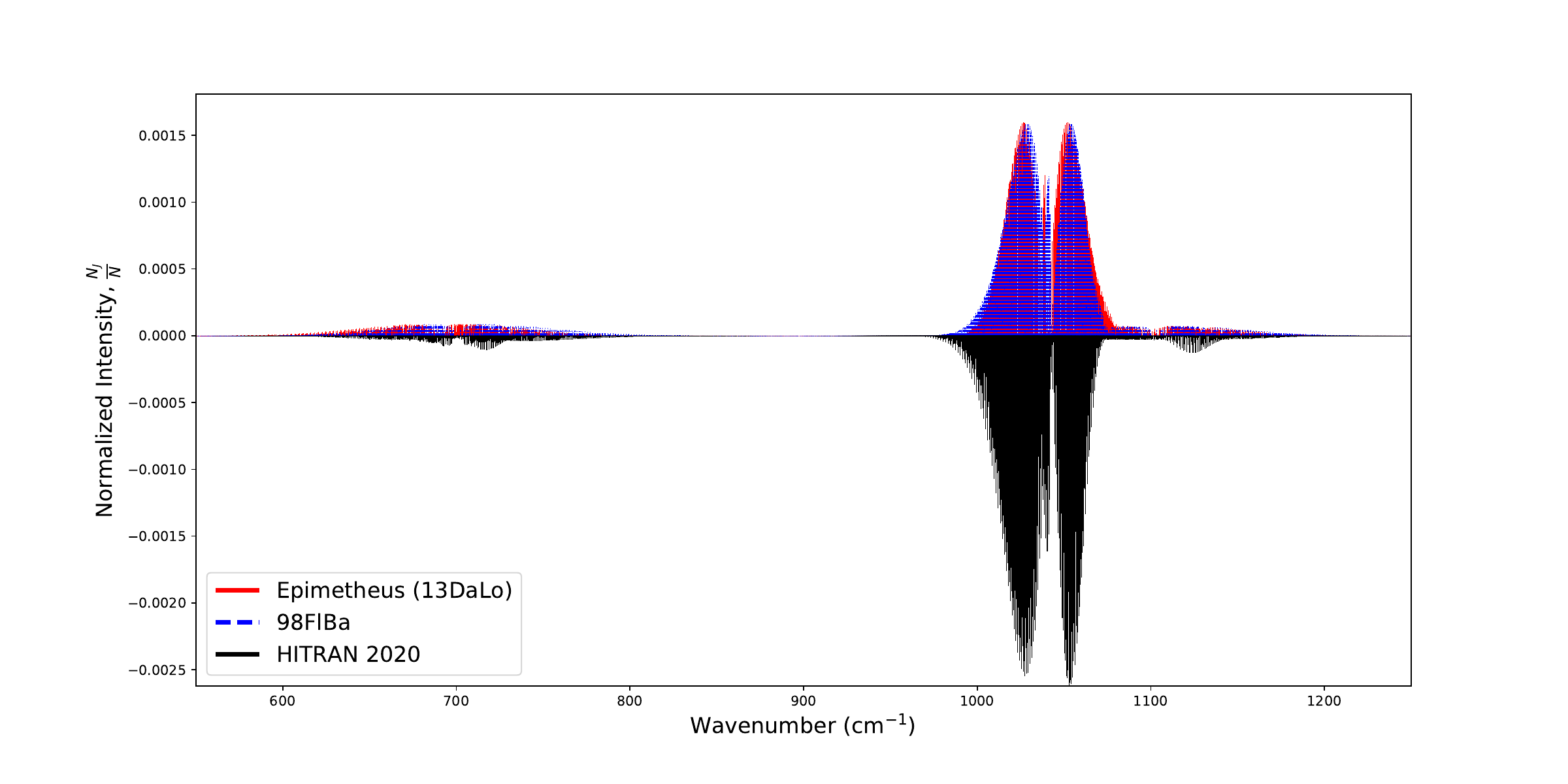}
    \caption{A comparison of the ozone 666 ($^{16}$O$_3$) spectrum from Epimetheus (13DaLo potential) in red. Data from the HITRAN 2020 release is in black and inverted for comparisons with our results. We have also included a theoretical best result from Epimetheus using literature constants from \citet{97FlBaXX} for additional comparisons in blue.}
    \label{fig:EPI_O3_666_13DaLo_98FlBa_HITRAN2020_300K}
\end{figure}


Figure \ref{fig:EPI_O3_666_13DaLo_18PoZo_HITRAN_subplots_300K} compares the spectra produced for the 13DaLo and 18PoZo PES with the HITRAN 2020 data by \citet{90FlCaRi, 22GoRoHa,03FlPiCa, 02WaBiSc, 21TyBaMi}. For this figure (and for all the following figures) HITRAN data are represented as inverted black spectra. The spectrum for 13DaLo reproduces the HITRAN spectrum for the asymmetric stretch mode (001, $\nu_3$) better than the 18PoZo PES, mainly due to a slight band origin shift for the latter. The bending mode (010, $\nu_2$) is better represented by 18PoZo, whereas neither models the symmetric stretching mode (100, $\nu_1$) accurately. Our spread of the transitions across the spectrum appears to be of similar range to that of HITRAN. 
Overall, both potentials offer a spectrum which qualitatively reproduces the main features of the HITRAN spectrum. 

Owing to a better quantitative match between HITRAN, and the spectrum obtained with the 13DaLo PES, particularly for the key mode 001 of the ozone spectrum,  we will mainly consider this PES for the remaining figures. It is worth noting that the discrepancy seen for the 18PoZo PES is likely caused by our approximate treatment of anharmonicity, rather than the quality of the 18PoZo PES itself which is not questioned.

In figure \ref{fig:EPI_O3_666_13DaLo_HARMONIC_HITRAN2020_300K}, we show that the harmonic approach generates a spectrum that is substantially displaced compared to Epimetheus' anharmonic approach. To the point that modes 100 and 001 in the harmonic approach could be easily confused. For simplicity, we only use for the comparison here the spectrum obtained with the 13DaLo PES. We also note that the harmonic lines for mode 100 are mostly swamped by the 001 mode, presumably due to the large anharmonic correction missing from the harmonic frequency for mode 001. We can see for the bending mode (010) that both methods (TOSH or harmonic) provide similar results, with the Epimetheus method over-correcting the band origin and the harmonic method naturally underestimating the band origin.

Since the Pandora approach does not account for all rovibrational effects, we test the accuracy of the TOSH (Epimetheus) spectrum by overlaying it with a Pandora spectra that uses the literature constants instead. Indeed, this test uses the same rovibrational simulation approach (Pandora) but highlights the effect caused by small changes of spectroscopic constants on the resulting spectrum. This is shown in figure \ref{fig:EPI_O3_666_13DaLo_98FlBa_HITRAN2020_300K}, where we used constants from \citet{97FlBaXX}, indicated by the abbreviation 98FlBa. We can see from the figure that the Epimetheus spectrum using the 13DaLo potential approximates the hypothetical best results of 98FlBa with high fidelity. The only major variations occur in the overall breadth of the R branch for mode 001 (we can see the red of 13DaLo on both sides of 98FlBa) and the band origin of the bending mode. The increased width of this branch is most likely due to the overestimation of the rotational constants, which leads to a wider spread of transitions when used within the branch equation (eq. \ref{eq:R_branch}). 

Based on the results shown in Figure \ref{fig:EPI_O3_666_13DaLo_98FlBa_HITRAN2020_300K}, Pandora can well replicate the HITRAN spectra. Some shortcomings occurs, however, especially with mode 100. The shape of this band is not replicated well and instead of having a single peak, our method produces a typical bimodal band. Another difference appears with the bending band which appears on the HITRAN spectrum to have more of a hybrid band structure (hybrid bands are discussed in more depth later in our results). Our spectral model in this case is limited to a single B vibrational band typing for the bending mode. 

Regarding the overall relative intensities of the bands, we see that our ratio differs slightly from that of HITRAN, most noticeably the 001 band not having as much weighting. This is something that is not calculated within our code, and relies on the reported literature values for the intensity ratios. In this instance, we have directly taken the ratio of 0.039 ($\nu_1$): 0.047 ($\nu_2$): 1 ($\nu_3$) from \citet{94IvPaXX}, which came with no errors. 

\subsection{Other isotopic variants}
\label{subsection:Results_668}

\begin{figure}[ht]
    \centering
    \includegraphics[width=\columnwidth]{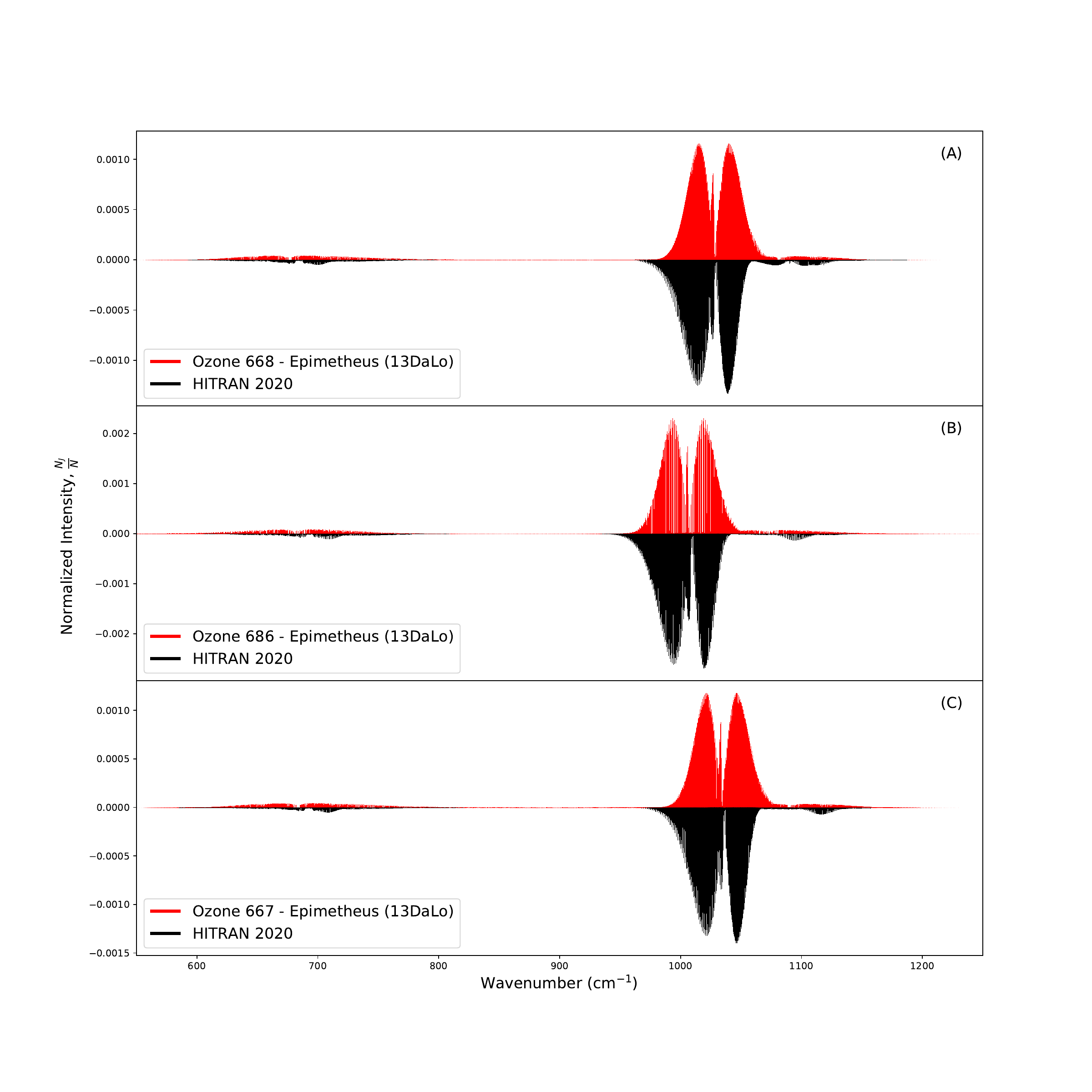}
    \caption{All Epimetheus spectra presented in this graph uses the constants derived from the \citet{13DaLoLi} potential. All graphs compare the Epimetheus spectrum against the data from the HITRAN 2020 release \citep[][]{22GoRoHa}. (A) The graph for ozone 668 ($^{16}$O$^{16}$O$^{18}$O). (B) The graph for ozone 686 ($^{16}$O$^{18}$O$^{16}$O). (C) The graph for ozone 667 ($^{16}$O$^{16}$O$^{17}$O).}
    \label{fig:EPI_O3_668_686_667_megamix}
\end{figure}

\begin{figure}[ht]
    \centering
    \includegraphics[width=\columnwidth]{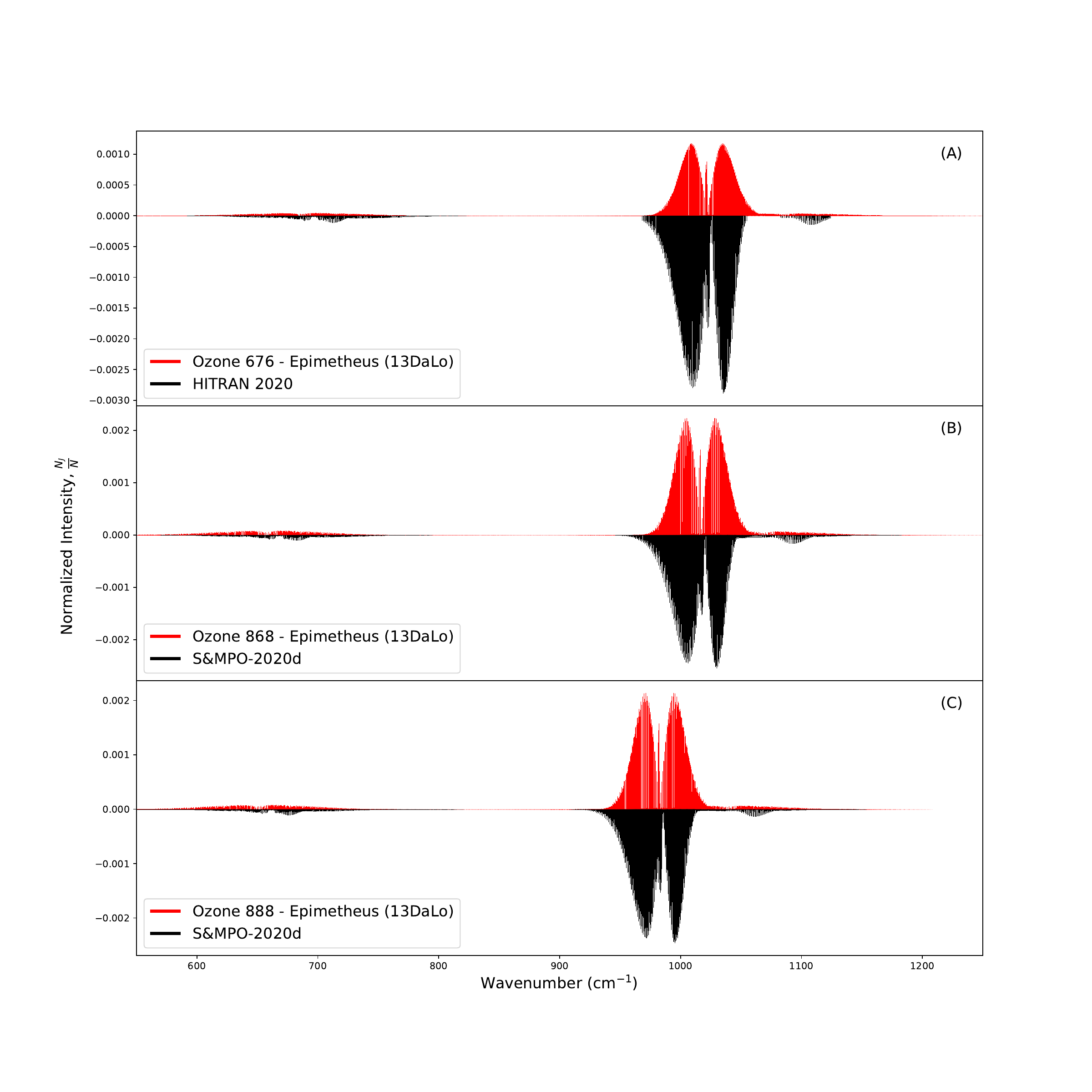}
    \caption{All Epimetheus spectrum presented in this graph uses the constants derived from the \citet{13DaLoLi} potential. Graph (A) compares the Epimetheus spectrum against the data from the HITRAN 2020 release \citep[][]{22GoRoHa}. Whereas graphs (B) and (C) use the data from the S\&MPO-2020d catalogue for comparison \citet{14BaMiBa}. (A) The graph for ozone 676 ($^{16}$O$^{17}$O$^{16}$O). (B) The graph for ozone 868 ($^{18}$O$^{16}$O$^{18}$O). (C) The graph for ozone 888 ($^{18}$O$^{18}$O$^{18}$O).}
    \label{fig:EPI_O3_676_868_888_megamix}
\end{figure}

We now investigate each of our remaining selected isotopomers, by comparing the spectra Pandora is able to produce (using the spectroscopic constants derived by Epimetheus using the 13DaLo potential) and the best experimental data available. The isotopomer spectra have been grouped together into Figures \ref{fig:EPI_O3_668_686_667_megamix} and \ref{fig:EPI_O3_676_868_888_megamix}, and each subplot is labelled to indicate which isotopomer it relates to. 

Ozone 668 is an asymmetric variant of ozone and therefore belongs to the C$_s$ symmetry group. \citet{97FlBaXX} suggested that for the asymmetric variants, all the vibration-rotation bands are hybrid bands which exhibit both A and B type components. In Pandora, we only model a single band for each mode and use instead the theory for the symmetric variants. We assume that the symmetric component is the dominant contribution to the hybrid bands, which seems justified in practice (Fig.\ \ref{fig:EPI_O3_668_686_667_megamix}). 

The spectrum for this isotopologue is compared to 668 HITRAN 2020 data by \cite{22GoRoHa, 90FlCaRi, 18BaStBa}, and is shown in figure \ref{fig:EPI_O3_668_686_667_megamix} (panel A). We see that our spectral modelling reproduces modes 010 (around 700~cm$^{-1}$) and 001 (around $1000-1050$~cm$^{-1}$) quite well, conversely mode 100 is not as well reproduced.

Our spectral model does not show the same characteristic PQR branch as the HITRAN2020 spectrum, specifically in replicating the small Q branch. This is presumably due to the omission of an additional A vibrational band, since we only model the B type for $\nu_1$. From a qualitative viewpoint the 100 mode for Epimetheus P branch is mostly swamped by the R branch for the 001 mode, something which does not occur for the HITRAN spectrum. This is due to the displacement in our band origin for Epimetheus from the observed frequency. In this particular scenario this is simply a minor issue, as the 100 mode contributes few substantial features to the spectrum.

Moving onto the next ozone isotopologue, 686, we know it is another symmetric ozone variant belonging to the symmetry group C$_{2v}$. Since this molecule is composed exclusively of $^{16}$O and $^{18}$O isotopes, spin effects will occur in the spectrum. As was the case for ozone 666, this means only levels of certain parity will be populated. The spectrum for this isotopologue is compared to 686 HITRAN 2020 data from sources \citep[][]{22GoRoHa, 03FlPiCa, 90FlCaRi} and is shown in figure \ref{fig:EPI_O3_668_686_667_megamix} (panel B).

The relative ratio of intensities between Epimetheus and HITRAN have a mixed accuracy for 686. The most apparent inaccuracy is for the 001 mode, where the maximum intensity for 686 is approximately half of that of HITRAN. On the other hand, the intensities of the bending mode 010 are well reproduced. While the rotational progression for mode 100 is very weak in our model, the overall shape of modes 010 and 001 agrees with the HITRAN spectrum, with it even displaying some of the same discontinuities that presumably arise from the spin effects.

The next isotopologue discussed is 667, which is an asymmetric variant of ozone and therefore belongs to the C$_s$ symmetry group. The spectrum for this isotopologue is compared to 667 HITRAN 2020 data \citep[][]{22GoRoHa, 03FlPiCa, 91RiSmDe}  and is shown in figure \ref{fig:EPI_O3_668_686_667_megamix} (panel C). 
Usually the presence of $^{17}$O isotope leads to a relaxation in population rules (something which has an important effect on the modelling and is discussed further for the 676 isotopomer), but this has no impact as 667 asymmetric species. This is because all the sub-levels are already populated with the same nuclear spin statistical weight as detailed in \citet{97FlBaXX}.

Our ozone 667 spectrum shows a good approximation of the HITRAN spectrum. Some issues with the branch widths are visible, in particular for mode 001 (which lies between approximately 1000-1090 cm$^{-1}$ with a wider the R branch. This is due to a slight over-estimation of the rotational constants, which prevents the expected rotational branch fold-back. As expected, the 100 mode shape is not reproduced well but the spread of transitions measure up with relative accuracy.

~\\
The first isotopologue shown on figure \ref{fig:EPI_O3_676_868_888_megamix} is ozone 676 (panel A). This isotopologue is compared to 676 HITRAN 2020 data \citep[][]{22GoRoHa, 87RoGaGo, 02WaBiSc, 91RiSmDe, 90FlCaRi, 85GaXXXX} and is another symmetric ozone variant (symmetry group C$_{2v}$). Since this isotopologue contains the $^{17}$O isotope, spin effects are different than if the composition was purely the $^{16}$O and/or $^{18}$O isotopes. As we can see in table \ref{tab:iso_abundance}, the spin for the $^{17}$O isotope is non-zero unlike $^{16}$O and $^{18}$O. For our approximate method, we have used the assumption that all sub-levels of any parity can be populated. However, depending on the parity (odd or even) of the asymmetric mode 001, different types of sub-levels gain a different nuclear spin degeneracy \citep[for details, see e.g.,][]{97FlBaXX}. 

Despite this, for 676 the 001 mode is reproduced quite well in the spectrum shown in the figure. The most notable difference would be the width of the P branch, due to the rotational constants not quite matching the target values. As expected from the results of the constants calculated in the previous Section, using the 13DaLo potential the origin of the 010 visibly appears to be shifted compared to the HITRAN reference spectrum. 

On the other hand, since we have assumed any sub-level can be occupied in 676, we observe an issue with the intensity ratios. This is most noticeable for mode 100, where the general shape of the branches are effectively flattened due to their low ratio value. We can also see that mode 010 is only just visible for the characteristic branch shapes and that mode 001 has a maximum intensity of roughly 0.0010, about 3 times lower than the reference HITRAN data. This is an issue as our spectra would be less prominent than expected for ozone, although still viable and is presumably a result of our assumptions regarding the sub-levels.

As the remaining isotopologues (868 and 888) are similar in composition and symmetry, we will discuss both together in this next section. 
The species 868 and 888 are shown in panel B and panel C of figure \ref{fig:EPI_O3_676_868_888_megamix}, respectively. Due to their compositions, only specific sub-levels are populated. We see that both simulated spectra provide a good approximation to the reference data. Some discrepancies arise for the 100 mode and there is a slight displacement, roughly 10 cm$^{-1}$, in band origins for 010. But overall Epimetheus and Pandora combined create spectra are of very good likeness to their respective reference data.

\section{Conclusions}
\label{section:Conclusions}

Based on our previous work on diatomics \citep{22CrBePi}, we present here the results of a new low-cost computational framework to determine the rotational constants of polyatomic molecules in their vibrational excited states. Our approach leverages and extends the existing TOSH theory from \citet{08LiGiGI} for anharmonic vibrational corrections so it can provide useful geometric corrections. Originally it was not presumed that this correction would be able to effectively translate from a diatomic theory to polyatomic, but generally we have seen from the data that we can approximate literature constants well.

Overall, we show that the results from our new code Epimetheus approximate the spectroscopic constants for ozone well. It was found that the potential used has a notable impact on the rotational constants calculated by Epimetheus. We have used two different potentials in our analysis, from \citet{18PoZoMi} and \citet{13DaLoLi}. We obtain that both potentials provide similar results if the species of ozone is symmetric, however once asymmetry arises \citet{13DaLoLi} struggles to match the accuracy of \citet{18PoZoMi} sometimes up to an order of magnitude.

We also have shown that the band origins, in particular for the symmetric modes, have discrepancies. The main cause of this issue is the known limitations of the TOSH method. It is due to the fact that the TOSH correction (in particular the $\sigma$ parameter), as described by \citet{08LiGiGI}, come from the truncated equations of the typical VPT2 theory and some of the constants, such as $\eta_{iii}$, become negligible due to the frameworks of their derivations. However, the strong trends within the anharmonically corrected band origin, discussed in our work, confirms the robustness of our code.

After having validated Epimetheus results using ozone, in forthcoming papers we will provide an extensive library of spectroscopic constants for a comprehensive list of polyatomic molecules. 

Additionally for this study, we have used our new code Pandora to generate molecular spectra. We show that in general our new simulation framework is capable of producing a good approximation of the desired spectra for the various isotopologues. In particular, our methodology produces better approximated spectra than that of the harmonic method, which is often quite inaccurate with respect to band origins and ultimately leads to incorrect branch shapes. This is expected, since its predicted rotational constants remain unchanged upon vibrational excitation. 

We have compared our spectral modelling results with the data from HITRAN 2020 and S\&MPO-2020d. The main deviations arising due to our method are the following: i) band origins sometimes have a noticeable displacement from the target spectra and ii) some of the details of the spectral features are not correctly predicted. Point (i) naturally arises from the band origin data from the TOSH framework and is not something we control. For point (ii), we have used a simple approximate spectral modelling method that utilises assumptions to enable low temporal and computational cost. Rather, full complex simulations which use high-resolution methodologies are required to obtain the detailed features of the molecular spectra. 

In order to estimate the impact of the assumptions made in the Pandora code, we have also used high accuracy experimental constants to produce molecular spectra; we still obtained some disparities with the benchmark data. This was expected because of the complexity associated in simulating the ozone molecules due to the properties associated with asymmetric top molecules. In a forthcoming study we are testing Pandora with different polyatomic molecules, to get a full assessment of the code performance. Among other upgrades, we will provide the capability to infer the vibrational band types using automation in the code rather than through manual selection as we did for the current work.
\\

\section*{Acknowledgements}

We acknowledge the support of JINA-CEE (NSF Grant PHY-1430152) and STFC (through the University of Hull’s Consolidated Grant ST/R000840/1), and ongoing access to {\tt viper}, the University of Hull High Performance Computing Facility. MP acknowledges support from the Lend\"ulet Program LP2023-10 of the Hungarian Academy of Sciences (Hungary), the ERC Consolidator Grant (Hungary) programme (RADIOSTAR, G.A. n. 724560), the NKFI via K-project 138031 (Hungary). This work was supported by the European Union’s Horizon 2020 research and innovation programme (ChETEC-INFRA -- Project no. 101008324), and the IReNA network supported by US NSF AccelNet.

%


\bibliographystyle{ApJ}
\bibliography{Main_body}



\appendix



\section{RCOM Workout}
\label{ap:RCOM_workout}
\begin{figure}[ht]
    \centering
    \includegraphics[width=0.2\columnwidth]{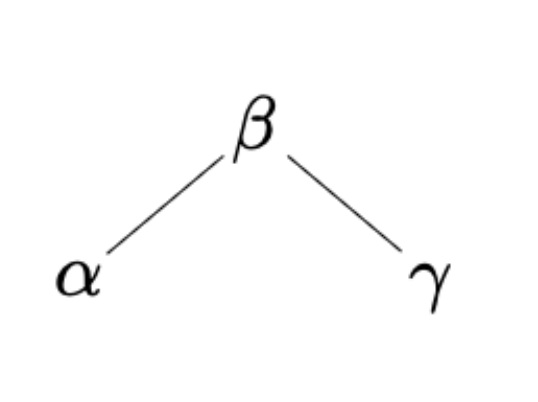}
    \caption{A 2D chemical structure depiction of imaginary molecule $\alpha$$\beta$$\gamma$.}
    \label{fig:chemfig_ABC}
\end{figure}

For this entire workout we use hypothetical molecule $\alpha$$\beta$$\gamma$ as the example. This has been done for ease of understanding. First we begin with the centre of mass equation, focusing on the x coordinate:
\begin{equation}
\label{eq:ABC_R_COM,x}
    \textbf{R}_{COM,x} = \frac{\left(m_{\alpha} \alpha_{x} + m_{\beta} \beta_{x} + m_\gamma \gamma_{x} \right)}{m_\alpha + m_\beta + m_\gamma}
\end{equation}
Where $\alpha_{x}$, $\beta_{x}$ and $C_{x}$ represents the x coordinates of the $\alpha$, $\beta$ and $\gamma$ atom respectively. We then continue this for the y and z coordinates.
\begin{equation}
\label{eq:ABC_R_COM,y}
    \textbf{R}_{COM,y} = \frac{\left(m_{\alpha} \alpha_{y} + m_{\beta} \beta_{y} + m_\gamma \gamma_{y} \right)}{m_{\alpha} + m_{\beta} + m_\gamma}
\end{equation}
\begin{equation}
\label{eq:ABC_R_COM,z}
    \textbf{R}_{COM,z} = \frac{\left(m_{\alpha} \alpha_{z} + m_\beta \beta_{z} + m_\gamma \gamma_{z} \right)}{m_{\alpha} + m_\beta + m_\gamma}
\end{equation}

We now compile the various $\textbf{R}_{COM,n}$ elements into an array:
\begin{equation}
\label{eq:ABC_R_COM,array}
    \textbf{R}_{COM} = 
\begin{bmatrix}
{R}_{COM,x} 
\\[6pt]
{R}_{COM,y} 
\\[6pt]
{R}_{COM,z} 
\end{bmatrix}
\end{equation}

To then calculate the position of an atom in the molecule (in this example we take the $\alpha$ atom), we take away the $\textbf{R}_{COM}$ positions from the origin positions: 
\begin{equation}
\label{eq:ABC_R_COM,A}
    \textbf{r}_{COM, \alpha} = 
\begin{bmatrix}
{\alpha}_{x} 
\\[6pt]
{\alpha}_{y} 
\\[6pt]
{\alpha}_{z} 
\end{bmatrix}
-
\begin{bmatrix}
{R}_{COM,x} 
\\[6pt]
{R}_{COM,y} 
\\[6pt]
{R}_{COM,z} 
\end{bmatrix}
\end{equation}

This ultimately gives us the COM position for the $\alpha$ atom. This process needs to be repeated for $\beta$ and $\gamma$.

\section{Inertia Tensor Components}
\label{ap:InertiaParts}

For completeness, we provide here the formulae employed for the various elements of the inertia tensor detailed in Section~\ref{subsection:Methodology_RevRotConst}, where $m_i$ represents the mass.
\begin{equation}
\label{eq:I_xx}
    I_{xx} = \sum_i m_i \left(y_i^2 + z_i^2\right)
\end{equation}
\begin{equation}
\label{eq:I_yy}
    I_{yy} = \sum_i m_i \left(x_i^2 + z_i^2\right)
\end{equation}
\begin{equation}
\label{eq:I_zz}
    I_{zz} = \sum_i m_i \left(x_i^2 + y_i^2\right)
\end{equation}
\begin{equation}
\label{eq:I_xy_yx}
    I_{xy} = I_{yx} = -\sum_i m_i x_i y_i
\end{equation}
\begin{equation}
\label{eq:I_xz_zx}
    I_{xz} = I_{zx} = -\sum_i m_i z_i y_i
\end{equation}
\begin{equation}
\label{eq:I_yz_zy}
    I_{yz} = I_{zy} = -\sum_i m_i y_i z_i
\end{equation}


\end{document}